\documentclass{article}
\usepackage{mathptmx} 
\usepackage{graphicx} 
\usepackage{subcaption}  
\usepackage[margin=2.5cm]{geometry}
\usepackage{authblk} 
\usepackage{amsmath} 
\usepackage{hyperref} 
\usepackage{cite}
\usepackage[numbers]{natbib} 

\title{Energy Consumption Optimization, Response Time Differences and Indicators in Cortical Working Memory Revealed by Nonequilibrium}

\author[1]{Xiaochen Wang} 
\author[1]{Yuxuan Wu} 
\author[2]{Feng Zhang} 
\author[3]{Jin Wang\thanks{Corresponding author: jin.wang.1@stonybrook.edu}} 
\affil[1]{College of Physics, Jilin University, Changchun, Jilin 130012, China} 
\affil[2]{State Key Laboratory of Electroanalytical Chemistry, Changchun Institute of Applied Chemistry, Chinese Academy of Sciences, Changchun, Jilin 130022, China} 
\affil[3]{Department of Chemistry and of Physics and Astronomy, State University of New York at Stony Brook, Stony Brook, New York 11794-3400, USA} 
\date{} 

\begin{document}

\maketitle
\begin{abstract}

The neocortex, a complex system governing multi-region interactions, remains a central enigma in neuroscience. Extensive research has produced quantitative insights into brain structure, ranging from microscopic to macroscopic scales. However, understanding the intricate mechanisms underlying neural activities remains a formidable challenge. From Hopfield network to recent developed large-scale cortical network, due to the rapid advance in connectomes laid the foundation for developments in neural or cortical network theory. While these theories capture cortical dynamics, they have limitations in representing global biological functions. In large-scale cortical models, an intriguing hierarchy of time scales has emerged, reflecting variations in information processing speeds among different brain spatial regions. The brain, a non-equilibrium living system, consumes significant energy costs, impacting evolutionary processes. Notably, long-distance connectivity between brain regions emerges as an evolutionary choice, suggesting a spatial organization on brain activity. To address these complex questions, we introduce a nonequilibrium landscape flux approach to assess large-scale cortical networks. We quantify potential landscapes and principal transition paths, unveiling inherent dynamical characteristics with varying timescales. We explore whether this temporal hierarchy is influenced by the spatial distribution of stimuli. Additionally, we investigate the intrinsic tendency of hierarchical networks to respond differently to stimuli. We extend our analysis to quantify the thermodynamic cost of sustaining cognitive processes, revealing a correlation with network connection patterns. Our analysis provides practical insights into energy consumption during cognitive processes, emphasizing the advantages of spatial distributed patterns in working memory tasks. Measuring certain characteristics experimentally can be challenging, and suitable comparisons may be scarce due to evolutionary changes. Therefore, our theoretical approach offers a valuable avenue for quantifying and predicting complex characteristics. By quantitatively assessing time irreversibility and critical slowdown, we gain insights into network bifurcations and state transitions, providing practical methods for the prediction and detection of state switching in cortical activities. These results enhanced our understanding of cortical activity.

\end{abstract}

\section*{Significance Statement}

In this work, we investigate the dynamics of cognitive processes emerging from complex interactions within cortical networks. Our findings reveal significant relationships between functional and structural properties, such as the energy-stability-flexibility trade-off and differences in time scales, which align with experimental observations and evolutionary trends. Additionally, this research bridges neurobiology with underlying mechanisms in statistical physics. The theoretical framework developed here is broadly applicable, offering insights into a wide range of biological systems.

\section{Introduction}

The neocortex represents an intricately complex system, orchestrating a multitude of multi-region interactions
\cite{RN1}. The fundamental inquiry into how cortical regions engage in such interactions and coalesce to give rise to observable behaviors has remained a paramount question in the field of neuroscience
\cite{RN1}. Over the years, a plethora of experimental neuroscience methodologies has emerged, such as the brain connectome and neurophysiology of behaving animals. These pioneering investigations have yielded a wealth of quantitative data, spanning the spectrum from microscopic to macroscopic architectures of the brain's structure
\cite{RN2,RN3,RN4,RN5,RN6}. 

Within the realm of neuroscience, understanding the intricate mechanisms governing neural activities is a complex endeavor
\cite{RN7}. While reductionist methods excel in recording the behavior of individual neurons or populations, a comprehensive bottom-up theory is required to elucidate the underlying physical mechanisms behind the observed behaviors
\cite{RN7}. John J. Hopfield proposed a physics-based model for neuron networks in the previous century, amalgamating neural encoding with statistical mechanics
\cite{RN8,RN9}. The Hopfield network simplified neurons as binary entities, akin to spins in physics, paving the way for subsequent developments on neural networks theory
\cite{RN10}. Afterwards, using self-consistent mean field theory and diffusion approximation to model interactions within neuronal populations, researchers have created a cortical network model, shedding light on the persistent cortical activities
\cite{RN11,RN12,RN13}. This model contributed to our understanding of functions as working memory, which rely on the sustained spiking of neuronal populations and have been explained using attractor models of dynamics. Subsequent work has extended this theoretical framework to encompass various cortical activities, spanning from local circuit dynamics to global cortical dynamics
\cite{RN26, RN21, RN15, RN25, RN16, RN17, RN19, RN22, RN27, RN20, RN23, RN18, RN1, RN14, RN24}. Recent advances in connectomes have led to the construction of large-scale cortical network models, offering insights into the distributed activity patterns underpinning cognitive processes as working memory and sensory decision-making
\cite{RN27, RN1}. These models provide a comprehensive view of the dynamics of cognitive processes on a multi-regional scale, revealing intrinsic nature such as robustness and flexibility. However, while these dynamic equations help to describe the cortical dynamics, they have limitations in fully representing the global characteristics relevant to biological functions. 

In the context of large-scale cortical models, due to the recent rapid development in connectomes, an intriguing hierarchy of time scales has been observed
\cite{RN21}. It is noted that primary sensory areas exhibit rapid responses, efficiently processing incoming information. Conversely, prefrontal and parietal cortical regions display slower transients characterized by ramping activity, a crucial element in the accumulation of information for decision-making processes
\cite{RN28, RN29}. This temporal differentiation can be attributed to the macroscopic gradient of biological features as delineated in mathematical analyses
\cite{RN21, RN27}. However, this insight alone does not suffice to elucidate why the original input stimulus tends to be directed towards lower hierarchy areas in the cognitive process. In other words, previous analyses had a prior assumption that the input stimulus primarily affects lower-level brain regions, and therefore, did not rule out the potential impact of spatial non-uniformity of stimuli on the response time scales
\cite{RN21, RN27}. If we adopt a weaker assumption, does a large-scale hierarchical network exhibit an intrinsic tendency to respond earlier in lower hierarchical areas and later in higher hierarchical areas? 

Being a non-equilibrium living system, the sustaining of a steady state in the brain requires dissipation and the flow of entropy, involving continuous energy exchange with the environment. It is well-established that the brain accounts for a significant proportion of the energy expenditure in mammalian organisms. Although the precise allocation of energy for cognitive tasks remains elusive, it is evident that the cost of cognition has evolutionary implications
\cite{RN30}. Beside the energy cost in every task, the general marginal costs also should include the cost on resource supply and transport for supporting the cognitive processes. So, the organisms have to pursuit better efficiency and face such tradeoff between function and cost
\cite{RN34, RN31, RN32, RN33, RN30, RN24}. Additionally, there is evidence that long-distance connectivity between brain regions appears to be an evolutionary choice, which may imply a spatially scaled influence in brain activity
\cite{RN35}. How does brain optimize the efficiency and achieve a balance of robustness, flexibility and energy consumption among the level of large-scale cortical networks? 

In this study, we introduce a comprehensive nonequilibrium landscape flux approach to assess the global dynamics of large-scale cortical networks
\cite{RN26, RN39, RN36, RN42, RN41, RN38, RN40, RN24, RN43, RN37}. We quantified the potential landscape and the principal transition pathways between network states.  Notably, these principal transition paths reveal spontaneous variations in timescales among hierarchical levels, illuminating the network's intrinsic dynamical characteristics. Additionally, we undertake the quantification of the thermodynamic cost associated with sustaining persistent activities, presenting a practical approach to gauge the energy consumption during cognitive processes. Comparing different network connection patterns, we uncover a correlation between thermodynamic costs and global or local connection strength, highlighting the potential advantages of distributed patterns in large-scale working memory tasks. 

Measuring certain characteristics, such as energy consumption, can be challenging through experimental means
\cite{RN34, RN31, RN32, RN33}. Moreover, suitable examples for comparison may be scarce due to their potential extinction during evolution
\cite{RN30}. Hence, our reliance on the nonequilibrium system approach offers a valuable avenue for quantification and prediction of complex characteristics through theoretical means. In the latter part of our study, we calculate several thermodynamic indicators when the network is subjected to sensory stimuli. Quantitative assessments of time irreversibility and critical slowdown provide insights and predictions regarding network bifurcations or transitions, aiding in the detection of state switching in cortical activities. As the attractor network models representing the cognition and other biological processes are significant and biological relevant, such analysis contributes to a broader understanding of the underlying biological processes not limited to the cognition. 

\section{Results and Discussion}
\subsection{Nonequilibrium Potential Landscape Quantifying Brain Network States and Transition}

The large-scale network model operates as a distributed working memory system with multistable dynamics. Attractors within the phase space represent different states of the brain network
\cite{RN45, RN44, RN15}. Bifurcations in the phase space denote transitions between the resting state and selective memory states, reflecting the decision-making and working memory processes 
\cite{RN26, RN46, RN27, RN23, RN14, RN11, RN24}. Compared to previous models that were based on local circuits, this large-scale model incorporates the contribution of multi-areas interactions to the cognition process, aligning more closely with recent evidence 
\cite{RN47} and offering greater potential for exploration. 

By applying a pulse input to specific populations in particular areas of the large-scale network, we observe a transition from a nonselective resting state to a selective memory state. For instance, a selective visual input target on the primary visual cortex (V1) causes the resting state to vanish, optimizing the related selective memory state and revealing the cognitive decision-making process
\cite{RN27}. Additionally, the selective populations' activity in certain areas (typically at higher hierarchy levels) is sustained even after transient pulse input or during reverse trials, indicating the persistence of working memory
\cite{RN48, RN27}. These processes are effectively demonstrated through the bifurcation diagram and the evolution of the probability potential landscape. Figure. 1a-b illustrates the bifurcation diagram concerning the stimulus strength input to V1 and the inactivation of target areas (9/46 v, 9/46d, F7, and 8B). 

\begin{figure}[h]
    
    \includegraphics[width=0.5\linewidth]{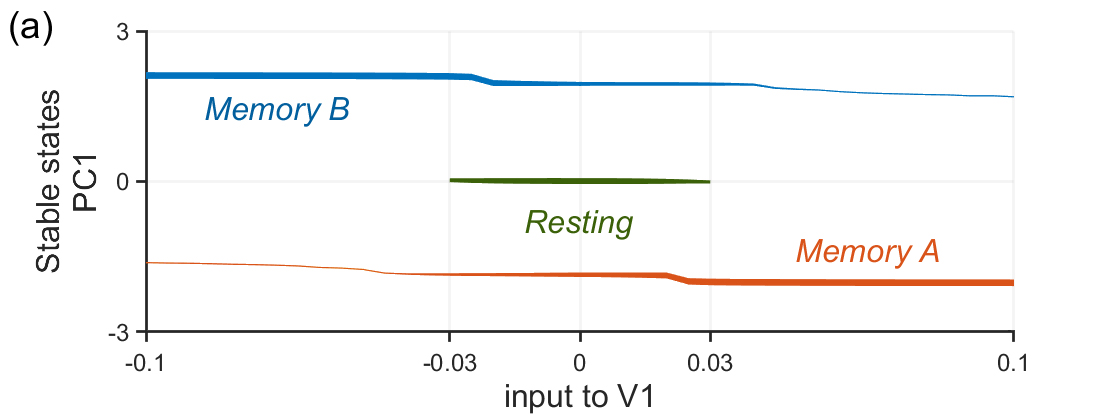}
    \includegraphics[width=0.5\linewidth]{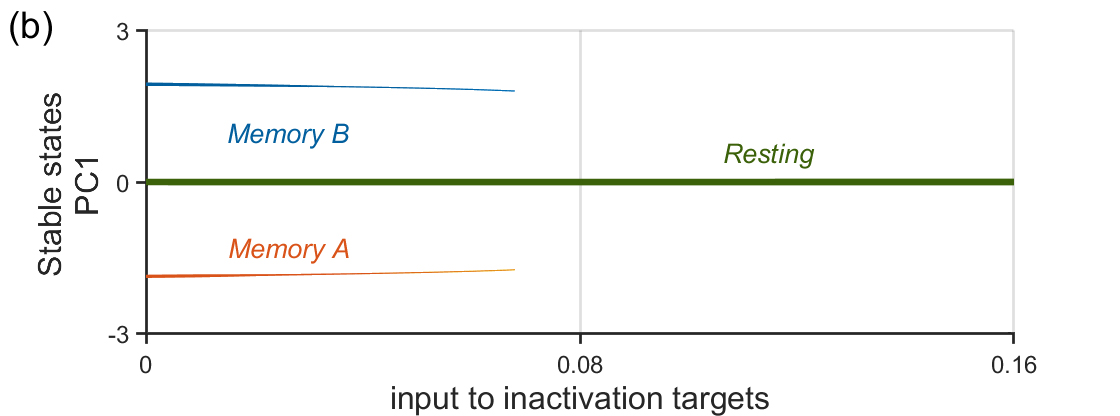}
    \includegraphics[width=0.33\linewidth]{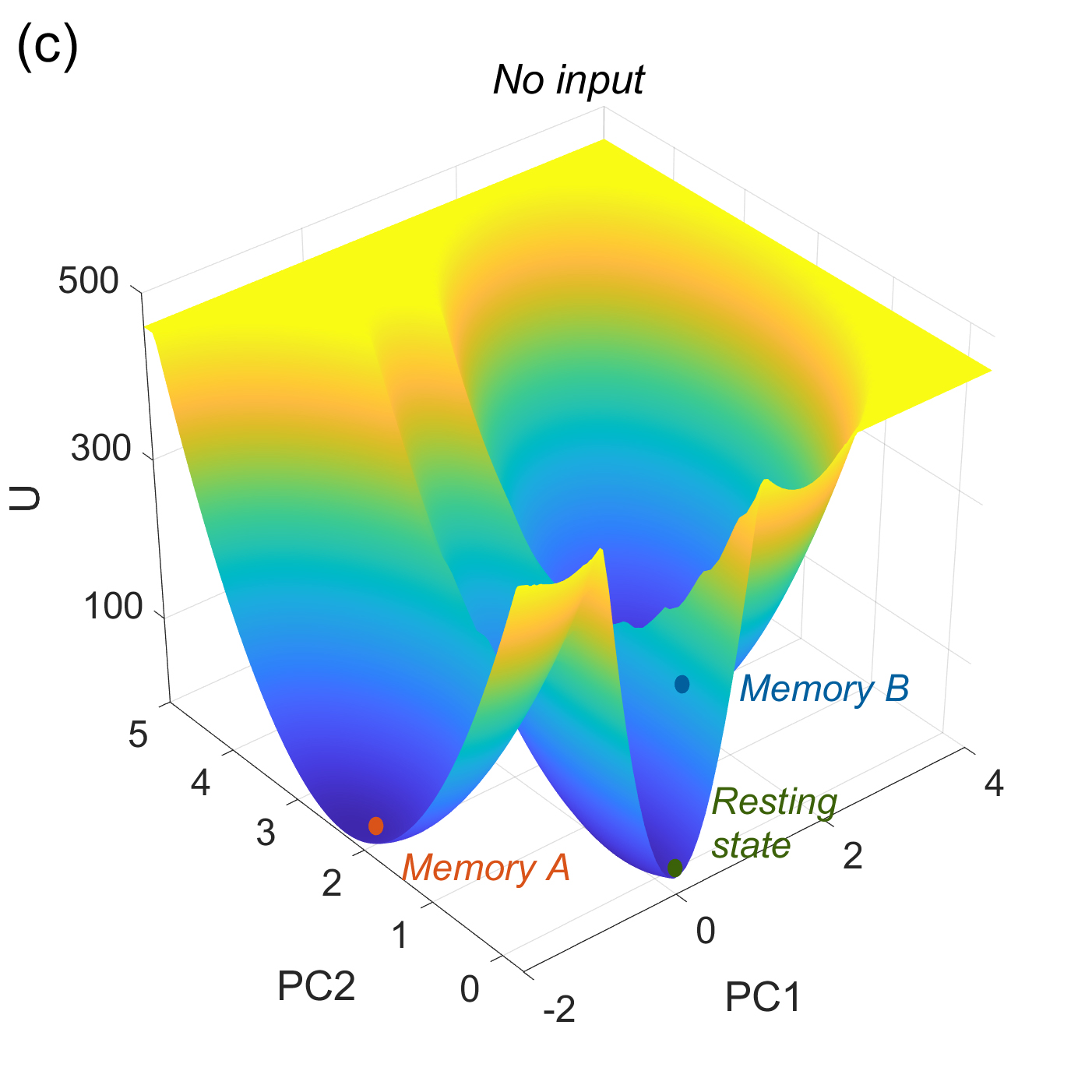}
    \includegraphics[width=0.33\linewidth]{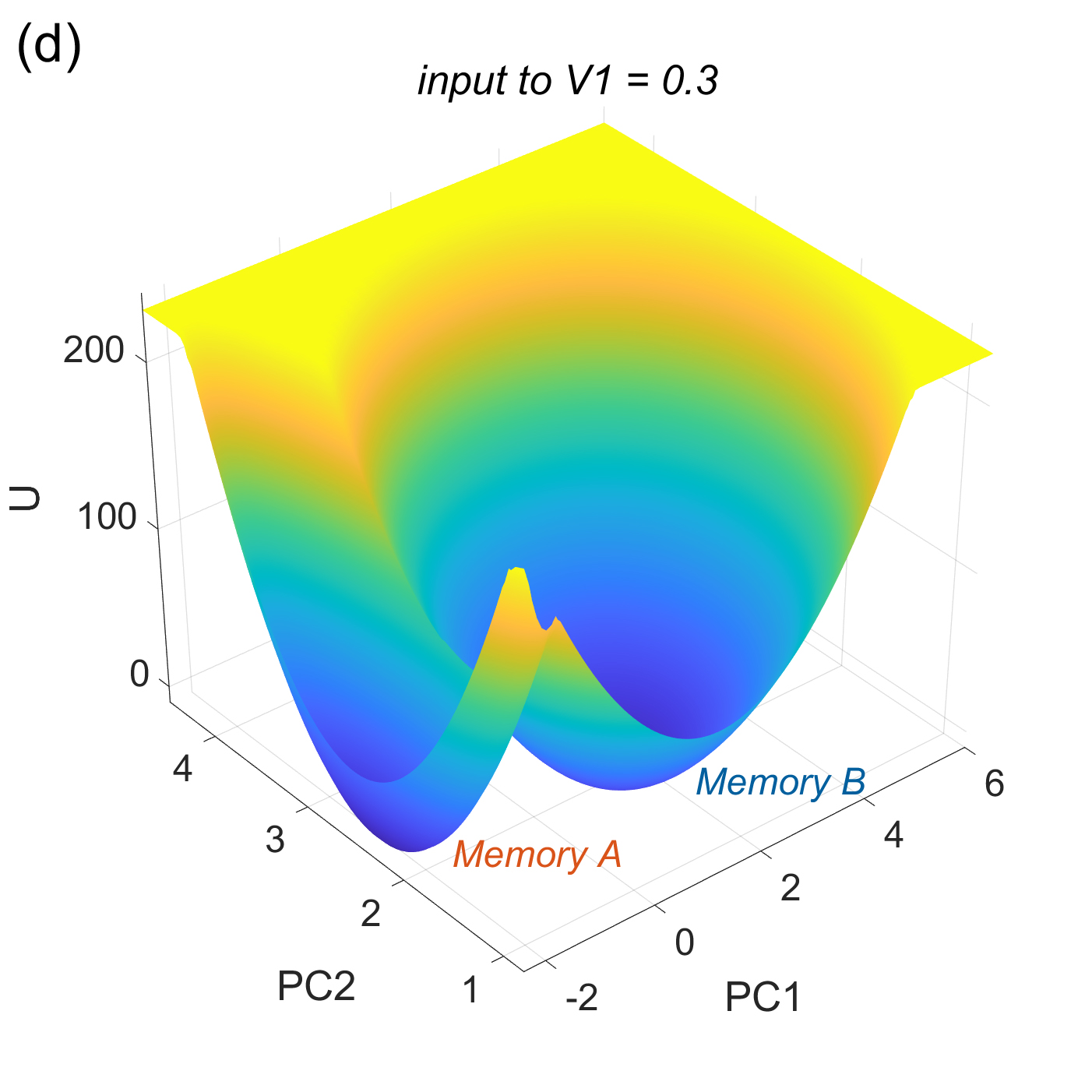}
    \includegraphics[width=0.33\linewidth]{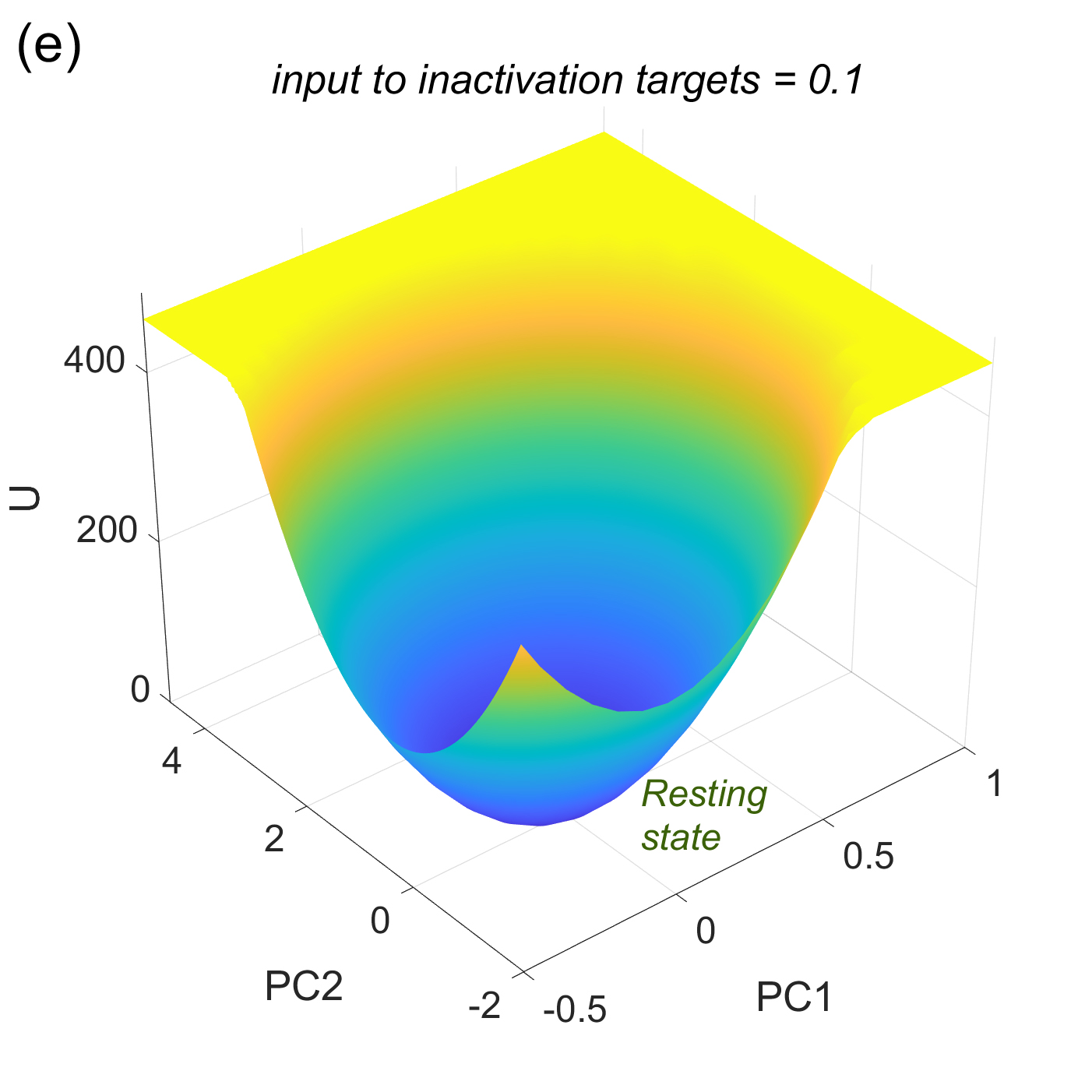}
    \caption{(a-b) The bifurcation diagram of large-scale network system with the activation stimulus strength to V1 and the inactivation stimulus to target cortical areas (9/46 v, 9/46d, F7, and 8B). The state coordinates are projected onto the first principal component (PC1). The bifurcation shows the scope of metastable. (c-e) The quantitative potential landscape for large-scale brain network under stimulus to certain cortical areas projected onto the subspace of the first two principal components (PC1 and PC2). We set the max synaptic connection strength constantly $J_{\max}$ = 0.3 in these plots.}
    \label{fig:enter-label}
\end{figure}

To quantify the potential landscape $U=-\ln  P_{\text{ss}}$ and the probability distribution, we solve the Fokker-Planck equation approximately using low temperature expansion and mean-field methods
\cite{RN49, RN50}. The three-dimensional potential landscapes ($U$) under finite fluctuations are shown in Figure.1c-e, projected onto the subspace of the first two principal components (PC1 and PC2). Since the original model comprises 90 dimensions, corresponding to the 90 populations in the large-scale network, we calculate the motions in this 90-dimensional space and identify the principal components based on the distribution of the state points. By projecting onto a 2-dimensional subspace, we visually reveal the potential landscape related to different steady states. 

In the absence of stimulus, three stable states exist in the phase space, including two memory states and one resting state (Figure. 1c). As the stimulus to the V1 area is introduced, the landscape evolves, gradually reducing the resting state until it disappears at the bifurcation point around $input$=0.035. With increased sensory stimulus, the related selective memory state becomes dominant, while the opposite memory state still persists, as illustrated by the topology of the landscape (Figure. 1d). 

The stimulus applied to primary sensory areas activates populations, introducing a state for working memory coding. Conversely, inactivating specific areas at particular hierarchy levels releases the memory state. For example, inhibiting a selective group of frontal areas (bottom row, including areas 9/46 v, 9/46d, F7, and 8B) with a shut-down input of particular strength provided to the nonselective populations of each area leads to a destabilized memory state
\cite{RN27}. The bifurcation emerges at $input$=0.083, where the memory states disappear, and above this strength, the network sustains a nonselective resting state. The potential landscape exhibits three basins under low inputs, illustrating two selective memory states and one nonselective resting state. As the input surpasses the bifurcation point ($input$ = 0.083), both memory states vanish, leaving only one basin for the resting state (Figure. 1e). 

\begin{figure}

    \begin{minipage}{0.36\textwidth}
        \includegraphics[width=\textwidth]{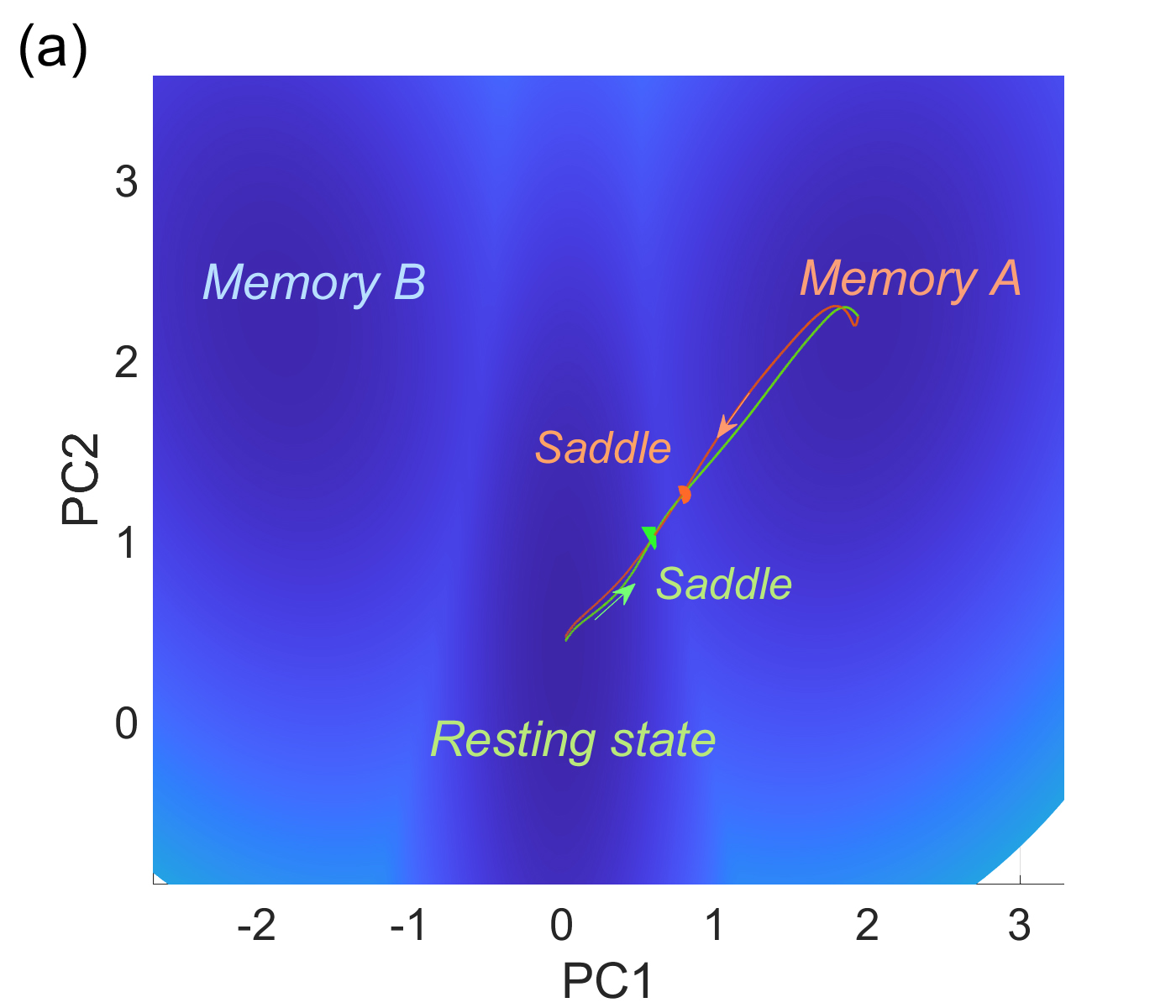}

        \vspace{0.0cm}  
        \includegraphics[width=\textwidth]{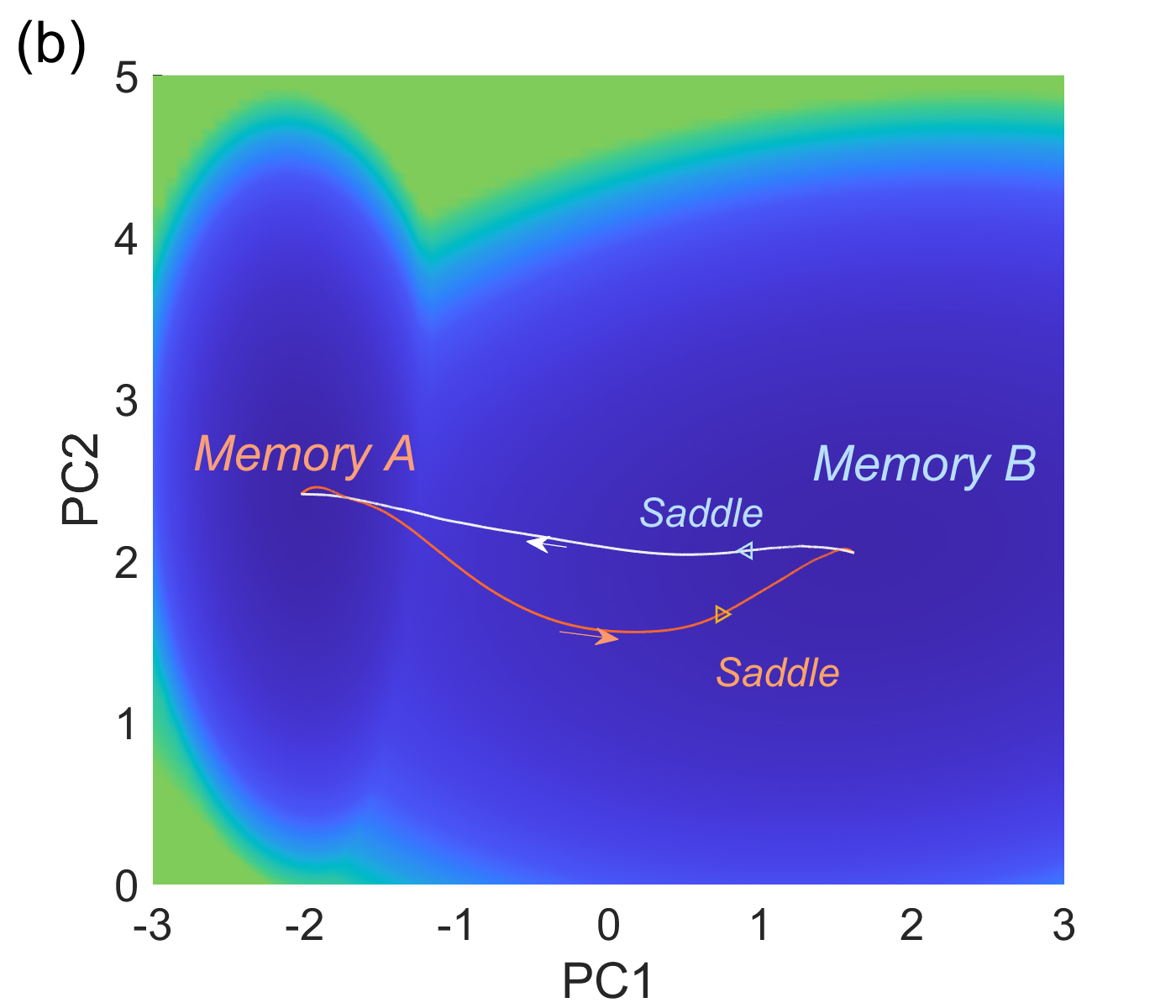}

    \end{minipage}%
    \begin{minipage}{0.64\textwidth}
        \includegraphics[width=\textwidth]{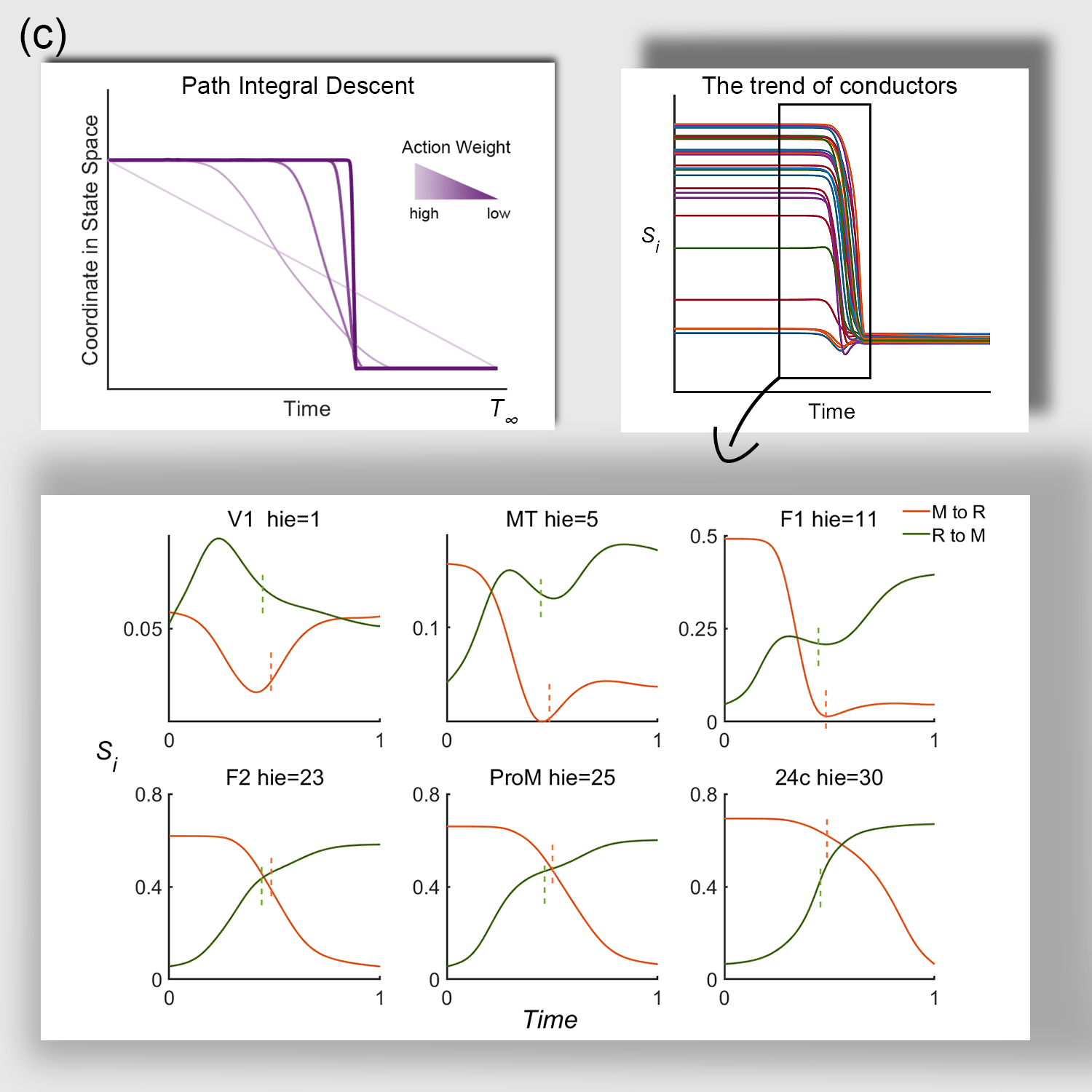}

    \end{minipage}



    \includegraphics[width=0.5\linewidth]{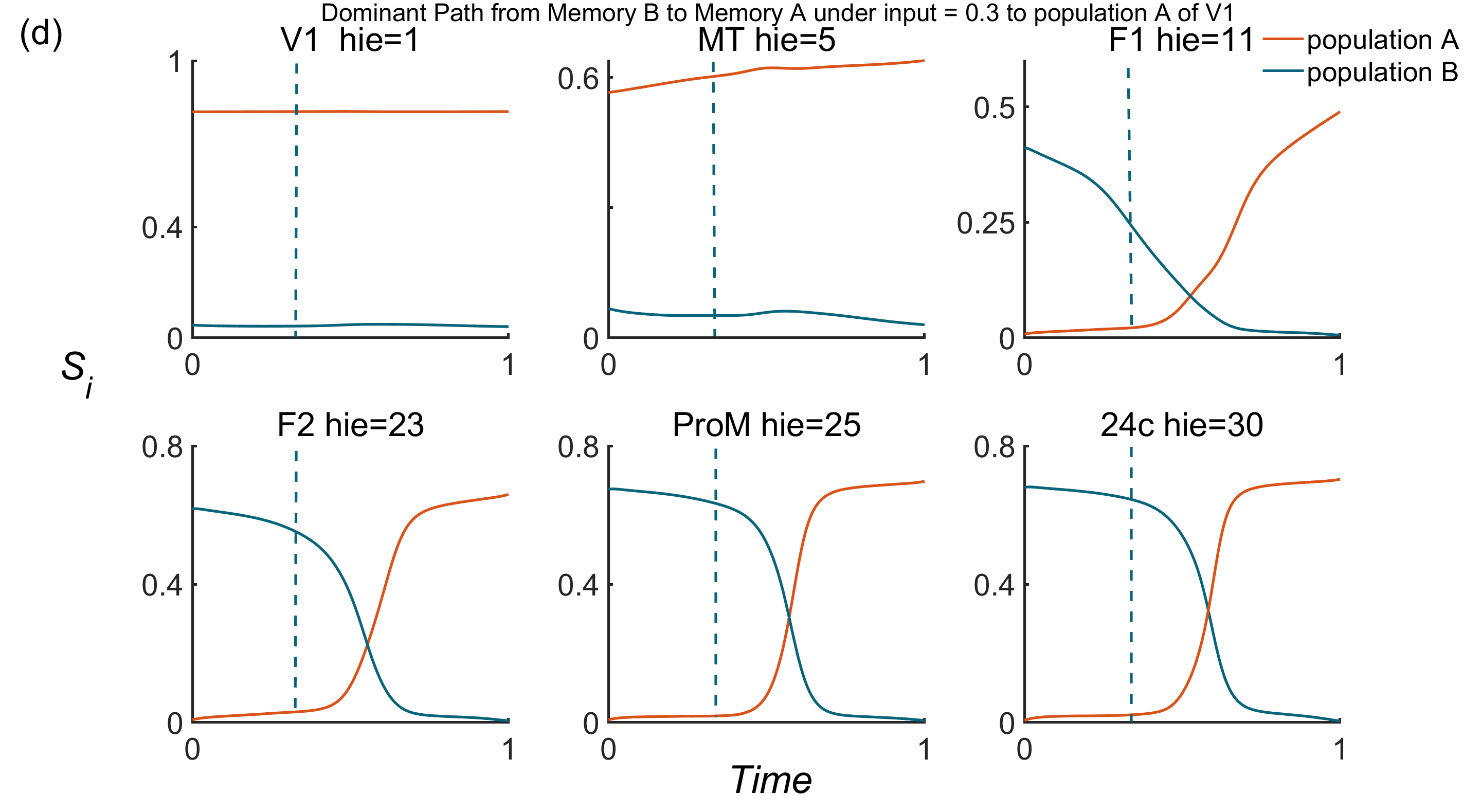}
    \includegraphics[width=0.5\linewidth]{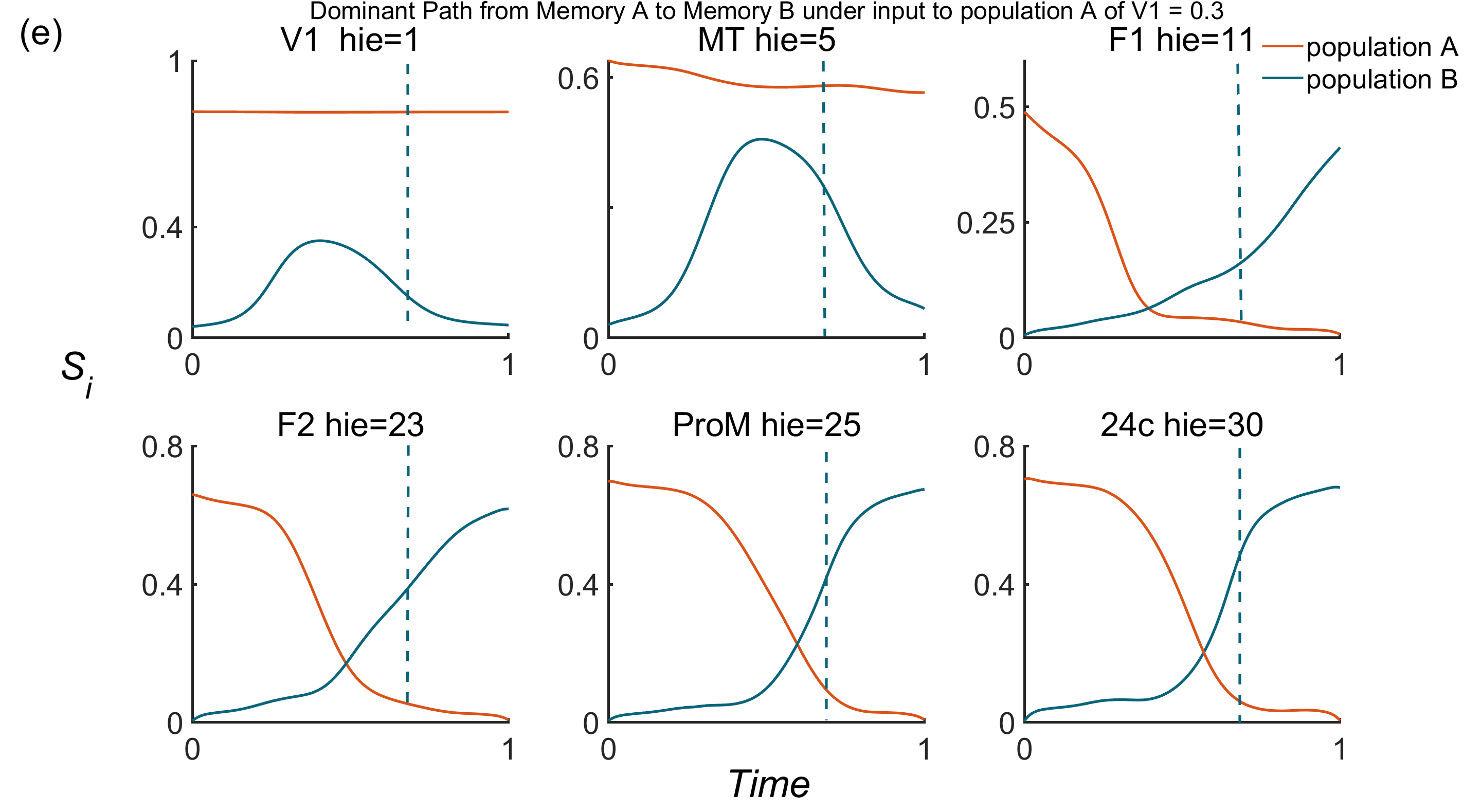}
    \caption{(a) The dominant path forward and backward between resting state and memory state. The saddles on both paths are illustrated here, which denotes the position of barrier strategic pass. The state coordinates are projected onto the first two principal components. (b) The dominant path forward and backward between two selective memory state under $input$ = 0.3 to population A of V1. The saddles on both paths are illustrated here, which denotes the position of barrier strategic pass. (c) Top-left: The optimization process of the transition path, where the color intensity indicates the descent in action, along with the evolving shape of the path. Top-right: The response trends of conductors ($S_i$), showing varying response speeds.Bottom: the trend of conductors among dominant path of related selective populations in several representative areas under no stimulus. The green line is the path from resting state to memory state and the red line is the path from memory state to resting state. The dash lines denote the position of saddle on both paths. (d-e) The trend of conductors among dominant path of related selective populations A (red line) and B (blue line) in several areas: (d) from B to A (e) from A to B. The dash lines denote the position of saddle on the paths. }
    \label{fig:enter-label}
\end{figure}

Additionally, we elucidate the principal transition path by optimizing the weight action derived from path integration and simulated annealing
\cite{RN51, RN39, RN55, RN54, RN53, RN52, RN38}. In Figure. 2a and 2b, we depict the principal transition paths that traverse forward and backward between the resting state and the memory states. Notably, the two prevailing paths exhibit disparities, underscoring the irreversibility attributed to the dominance of pathways due to the presence of nonequilibrium rotational flux. Concurrently, the distinction between the forward and backward paths finds explication in the activities within specific population areas, partially illustrated in Figure. 2c-e. Figure 2c illustrates the optimization process of the paths. In the top-left image, we observe how the shape of the path evolves as the path integral gradually descends. The principal path (depicted as the darkest line) corresponds to the trajectory with the lowest action. Over a long timescale, the principal path exhibits a sharp transition after a prolonged period of hesitation, which is characteristic of state transitions in multistable dynamics. However, when we zoom in on the sharp transition at a faster timescale, we notice different behavior.

\subsection{Temporal Hierarchy and Spatial Organization}

The difference of response time during the sharp state transition can be observed, which relate to the hierarchy of cortical areas. In the top-right image of Figure 2c, the sharp transitions of conductors ($S_i$) are shown. These conductors exhibit varying response speeds during the transition, corresponding to different cortical areas. The bottom image of Figure 2c highlights several representative cortical areas involved in this process. During the transition from the resting state to the memory state, the journey initiates with escalating spiking rates in the primary visual areas. Conversely, the higher hierarchy regions exhibit a gradual ascent until they reach a pivotal point. Subsequently, post the pivotal point, the lower hierarchy regions such as V1 and MT commence deactivation, concomitant with rapid escalation in spiking rates within the higher hierarchy areas. In contrast, while transitioning from the memory state to the resting state, the higher hierarchy regions (e.g., 24c) tend to sustain heightened spiking rates until the pivotal point, beyond which they experience a gradual decline. Meanwhile, areas possessing lower hierarchy but persistent activity (e.g., F1) exhibit significant deactivation before reaching the pivotal point. Notably, under visual stimulus (applied to V1), the system demonstrates two stable memory states, as illustrated in Figure. 1. These extant memory states may counteract external signals, showcasing the vital attributes of neural circuit systems – robustness and adaptability in response to environmental fluctuations. We proceed to compute the prevailing pathway between two discerning memory states under the influence of an external stimulus applied to population A within the V1 cortex (input = 0.3). Furthermore, the identification of saddle points along both trajectories enhances our understanding. Our findings illuminate that during the adaptive transition under the environment, the initial shift unfolds within the lower hierarchy cortical regions. Post surpassing the critical threshold, a rapid transformation occurs within the higher hierarchy areas. Conversely, when individuals make decisions in defiance of the environment, activation initiates within the primary regions, accumulating progressively before triggering a corresponding excitation within the higher hierarchy cortical regions. Many records substantiate the assertion that these cognitive task processes are under the governance of distributed patterns within the brain network 
\cite{RN15, RN47}. The optimized pathway reveals a bias in cortical activity distribution, both in temporal and spatial domains. This observation augments the existing evidence for distributed patterns in cognitive tasks. Moreover, these results intimate an ordered emergence of decisions across varying periods within the network hierarchy, notably highlighted by the differing cortical area intersections of the red and green lines in Figure. 2c. This intriguing behavior resonates with prior evidence 
\cite{RN21, RN20}. Furthermore, these findings diverge from prior evidence suggesting the utilization of a logically weaker premise assumption, thereby unveiling that the temporal scale disparity is an inherent attribute, rather than a consequence of the stimulus input area. 

\subsection{Thermodynamic Cost and Insights on Energy-function Trade Off}

The brain functions as an open system, engaging in the exchange of materials, information, and energy with the natural environment. According to the principles of nonequilibrium thermodynamics, the dynamics of the system is determined by both the gradient of the nonequilibrium landscape and the curl probability flux
\cite{RN38, RN56}. In cases of steady-state systems, the curl probability flux can be non-zero, signifying detailed balance breaking and a shift away from equilibrium. This curl flux serves as a measure of nonequilibrium and the net input or output of the system. Consequently, unlike equilibrium systems, nonequilibrium systems must sustain their activities through energy consumption or particle flux
\cite{RN57}, which is closely linked to the non-zero curl flux, denoted as $J_{ss}$. The associated thermodynamic cost can be quantified by the entropy production rate, represented as $\dot{S}=\int J\cdot D^{-1}\cdot J / P \, dx$, where $D$ signifies the diffusion coefficient, indicating the magnitude of fluctuations.

\begin{figure}
    \includegraphics[width=0.333\linewidth]{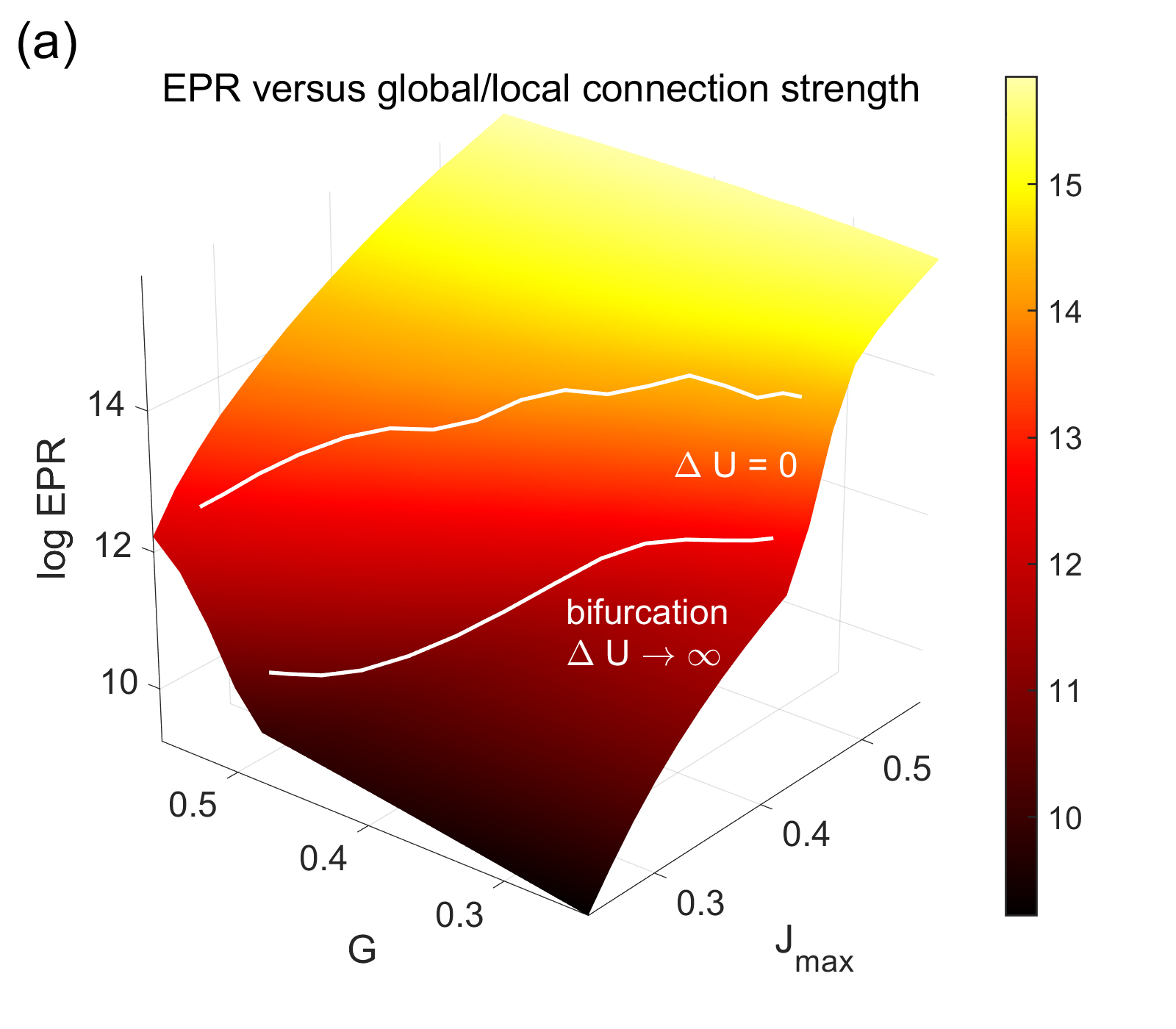}
    \includegraphics[width=0.333\linewidth]{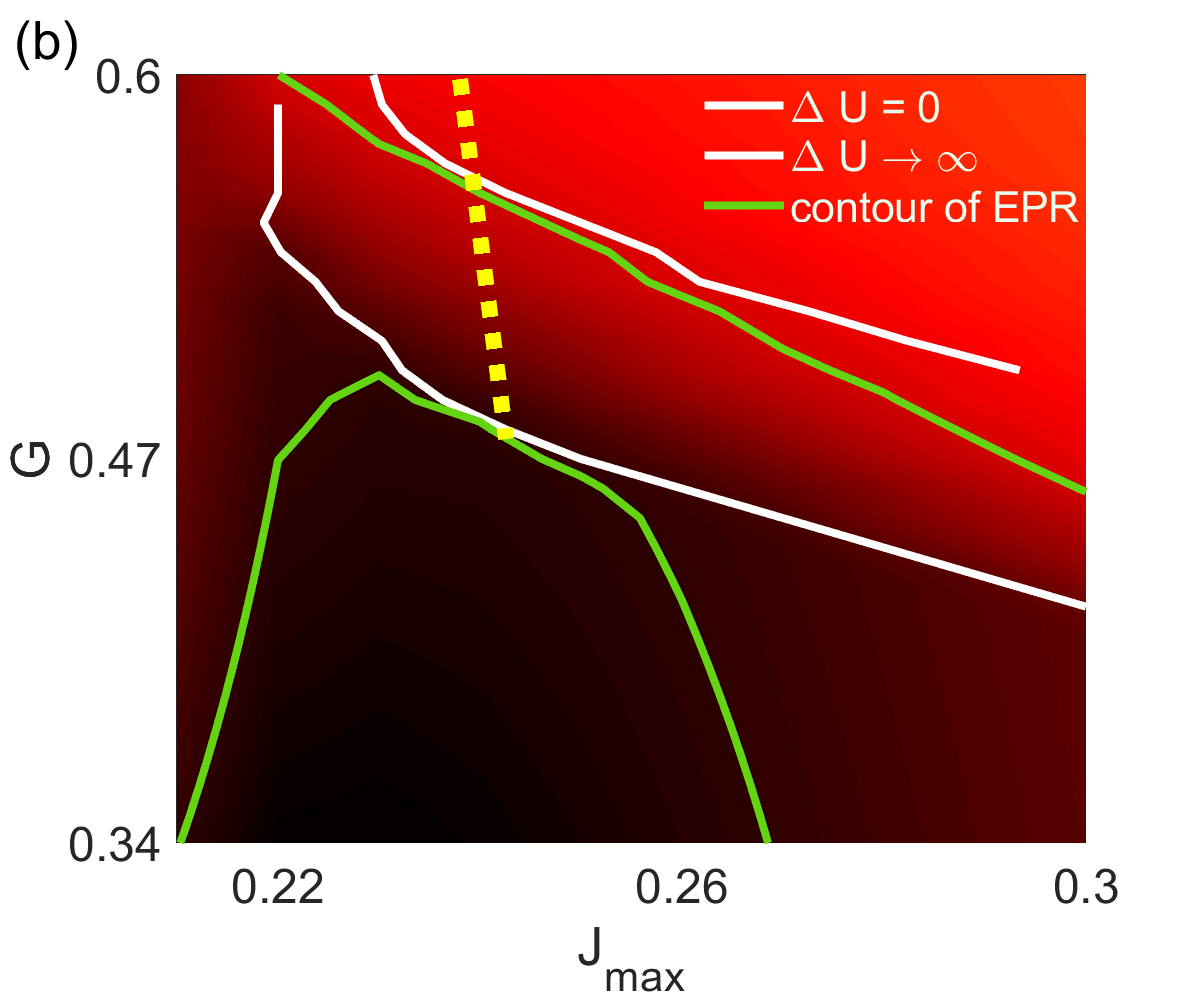}
    \includegraphics[width=0.333\linewidth]{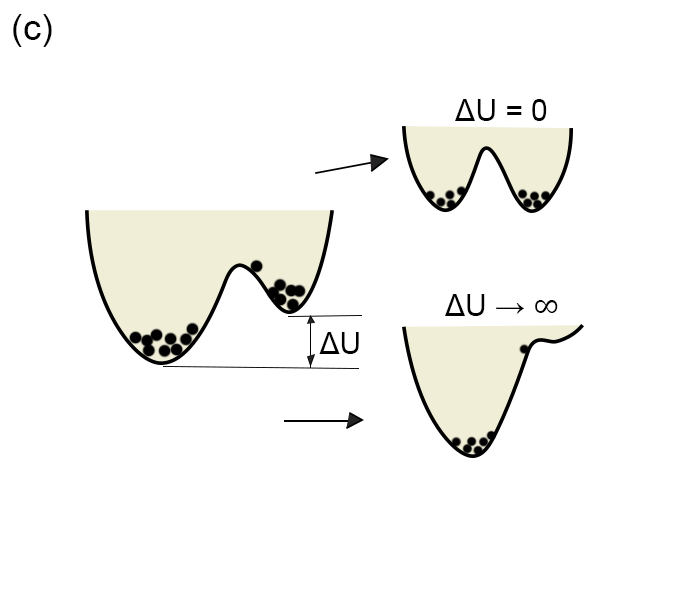}
    \includegraphics[width=0.333\linewidth]{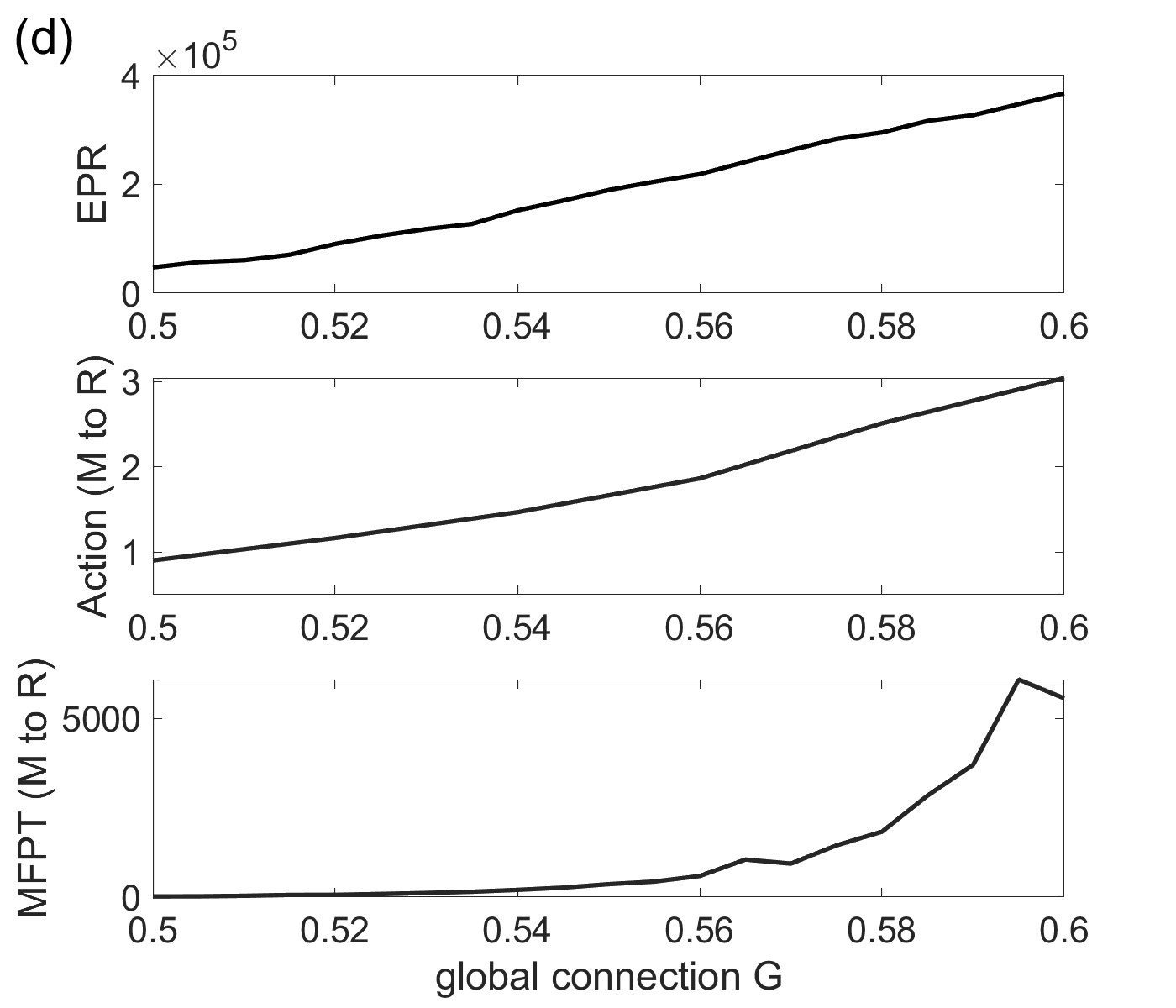}
    \includegraphics[width=0.333\linewidth]{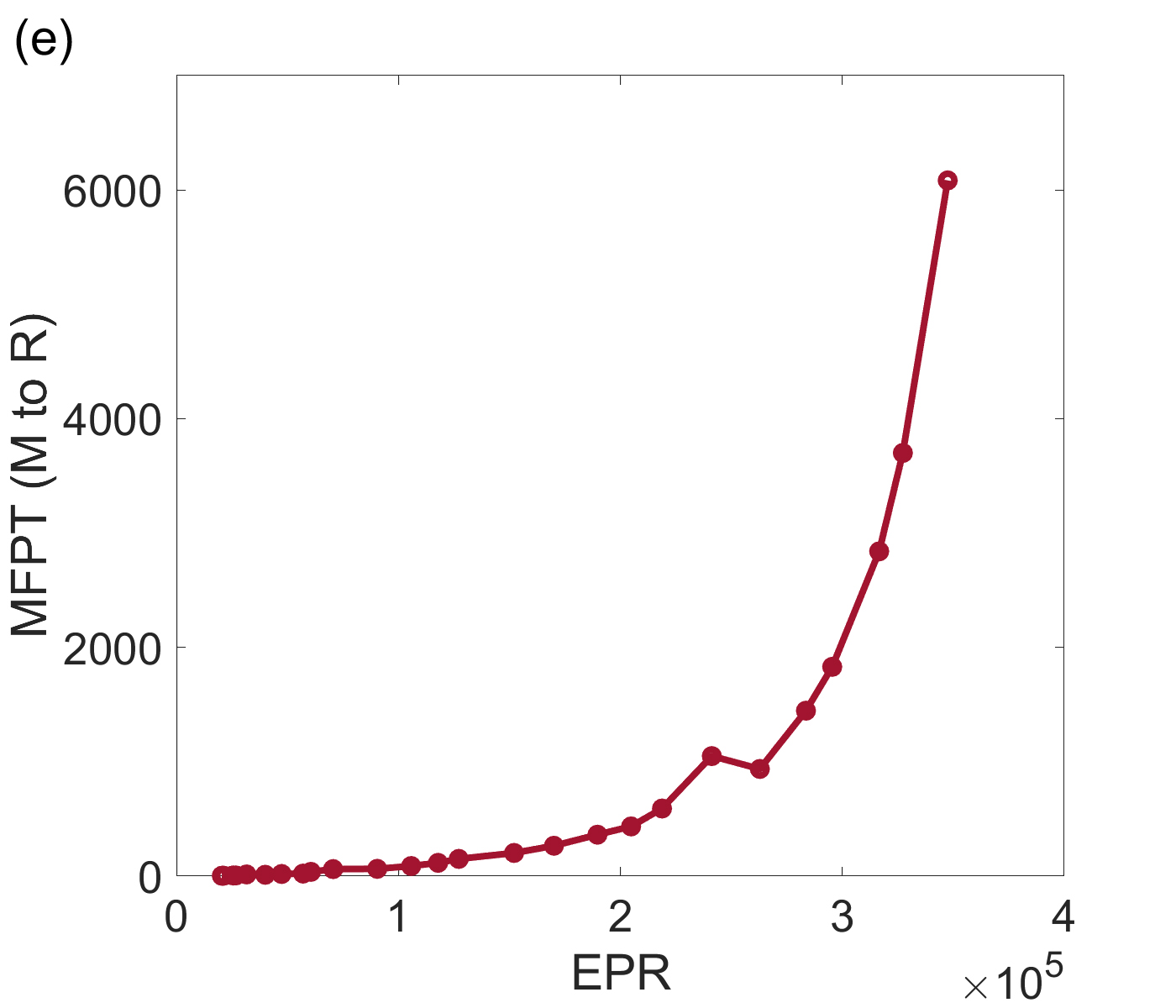}
    \includegraphics[width=0.333\linewidth]{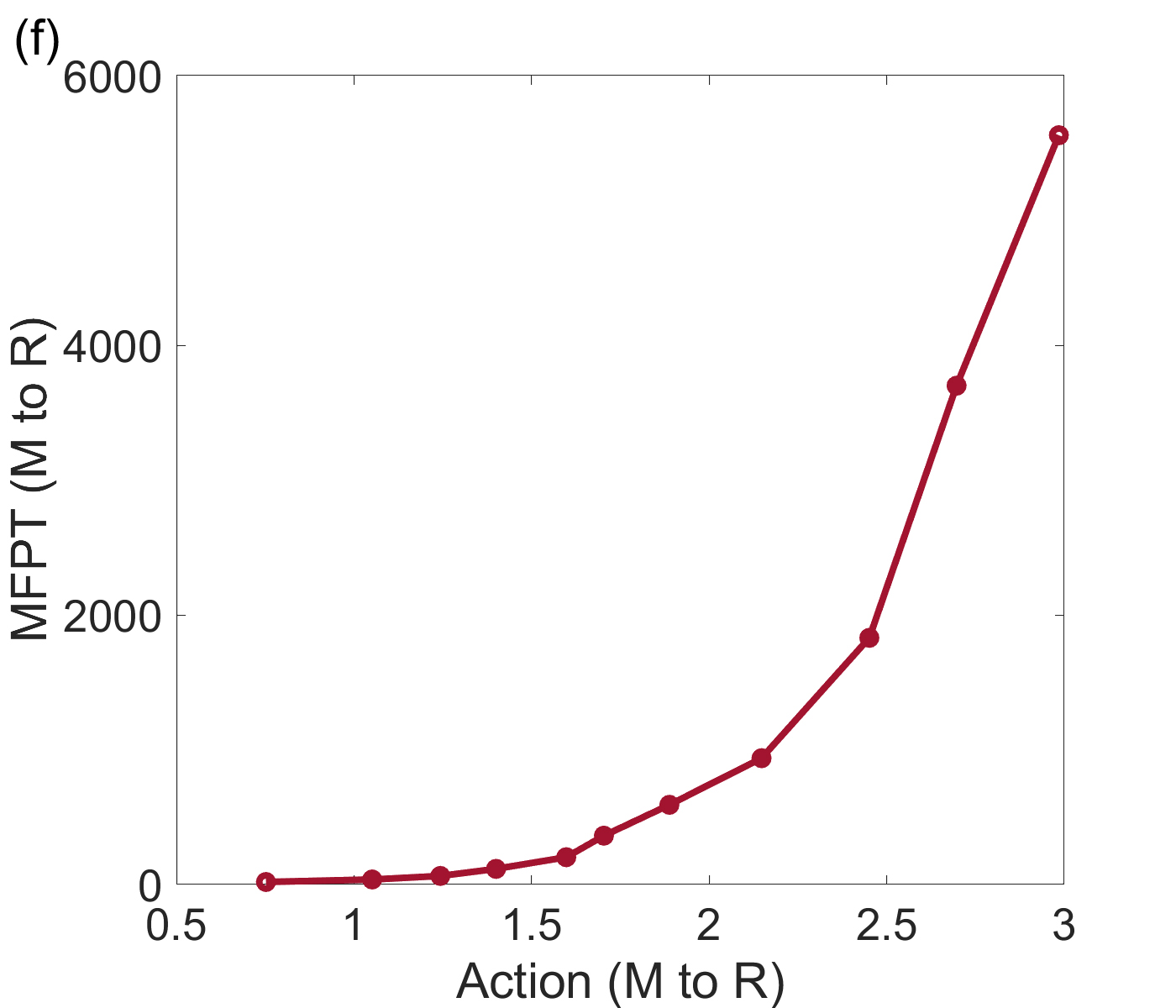}
    \caption{(a) Entropy Production Rate vs. Global Connection Strength ($G$) and Local Connection Strength ($J_{\max}$). The logarithmic trend of EPR reveals a significant increase with the enhancement of both global and local connections. The critical threshold, where memory states emerge or the barrier height ($\Delta U$) between the resting state and memory state becomes zero, is depicted by the white lines. (b) The indifference curve for EPR and ($\Delta U$) under different $G$ and ($J_{\max}$). The green lines are isolines of EPR. The white lines are isolines of ($\Delta U$). The yellow dashed line connecting the two intersection points indicates the optimal balance between $G$ and $J_{\max}$. (c) Illustration of the definition of relative stability ($\Delta U$) between resting state and memory states. $\Delta U$ represents the height difference between the bottoms of two basins, as previously illustrated in Fig. 1c. (d) The trend of EPR, Action, Mean First Passage Time vs. global connection $G$, when setting $J_{\max}$ = 0.243. (e) The relation between MFPT and EPR from the same data in (d). (f) The relation between MFPT and Action from the same data in (d). }
    \label{fig:enter-label}
\end{figure}

In Figure 3(a), higher connection strength (namely larger $G$ and $J_{\max}$
\cite{RN58} in the large-scale network) is associated with a significant increase of entropy production rate (EPR). As we know, higher EPR leads to higher energy consumption during performing cognition tasks, therefore enhancing the cognitive performance
\cite{RN61, RN60, RN24, RN59}. So, how is the cognitive performance related to the specific structure quantification? And how does the brain make the choice facing the trade-off between energy cost and performance? 

As for the performance of working memory, an important point we concern about is the duality of the memory traces. For a robust working memory function, the network should sensitively capture the information input and encode the information with a stable pattern. Here, to assess the network's robustness, we evaluate the topography of the landscape. As illustrated in Figure 1, the landscape's basins represent states in phase space, and the gradient of the landscape tends to stabilize these states. Consequently, the network's robustness, specifically the stability of the attractors, is closely tied to the topography of the landscape. We employ the relative stability by measuring the difference of the potential $\text{$\Delta $U}$ between the resting state and memory states as an indicator (Figure 3c). Notice that although the relative stability does not usually pin down the global network stability, the specifically selected $\text{$\Delta $U}$ can serve as an indicator for comparing the topography of the landscape. In the vicinity of the bifurcation point (indicated by the white line in Figure 3a), $\text{$\Delta $U}$ approaches infinity, leading to the emergence or disappearance of memory states. As $\text{$\Delta $U}$ decreases, memory states become more stable, with $\text{$\Delta $U}$ = 0 signifying the equal probability for both resting and memory states, and indicating the thermodynamic phase transition point
\cite{RN38}. 

Figure 3a shows that EPR sharply increases when system undergoes the bifurcation (marked as white line, {$\Delta U \rightarrow {\infty}$}), which denotes the emergence of memory states. Particularly, on the bifurcation line (white line), EPR is significantly declined when G increased. As the $\text{$\Delta $U}$ remains the same along the bifurcation line, the stabilities of the memory states are similar. Thus, a higher fraction of global connections can significantly reduce the entropy production rate over an order of magnitude while maintaining stability. 

For a more in-depth analysis, we find an optimal solution of the energy-function trade-off in theory. The systemic impacts of robustness and entropy production are integrated over time during delays. Consequently, the sustaining of stability carries a direct marginal cost, namely energy consumption, which is reflected in the entropy production. This trade-off becomes more significant as a longer delay time required, owing to the evident positive correlation between EPR (green lines) and relative stability (white lines). These facts allude the existence of an most economical pattern for this network model. In Figure 3b, we present indifference curves that illustrate where the EPR and relative stability remain the same levels under different connection strength. Because EPR and relative stability grow monotonically perpendicular to the indifference curves, the minimum EPR value on the white line should be at the tangent point (of green line and white line in Figure 3b). Therefore, the tangent points indicate where the maximum ratio of relative stability and entropy production is maintained, namely local optimal energy consumption choices. If we aim to optimize system properties by altering connection patterns, finding a balanced solution near the yellow dashed line connecting the two tangent points may prove beneficial. 

Consequently, the structure parameters have certain effects on the theoretical performance of cognitive function, especially the global connection strength. From Figure 3b, it can be observed that the tangent points are at relatively high $G$ and low $J_{\max }$, which implies the optimal pattern for energy efficiency. So, this result indicates a practical benefit of enhancing global connections within a network. 

As previously mentioned, relative stability cannot be used to exactly predict the absolute stability. So, we calculate the action by path integral from memory state to resting state, indicating the capability of communication between the states. To illustrate the relationship between EPR and stability within the optimal connection pattern, we maintained a fixed value of $J_{\max }$ at 0.243 for simplicity while adjusting the global connection strength parameter, $G$. In Figure 3d, we observe an upward trend in EPR, action, and mean first passage time (MFPT) as we increase global connection strength, $G$. MFPT measures the time required for a system, initially proximate to a memory state, to switch to another state. To enhance stability measurements' precision, we introduced additional Gaussian white noise into the dynamical equations during MFPT assessment. This noise accelerates the diffusion process without altering bifurcation behavior. Figure 3e and Figure 3f present the same data, emphasizing the relationships between MFPT, EPR and action, highlighting the connection between stability and energy cost. 

While direct experimental comparisons of energy costs may pose challenges, insights from evolutionary trends offer valuable clues for implementing these theoretical insights onto real biological systems. Evidence supporting this direction comes from studies on the spatial scales of cortical connections in mammalian brain regions
\cite{RN35}. These investigations characterized the statistical properties of interareal connections within the marmoset cerebral cortex, revealing a spatial embedding with a characteristic length. As the brain structures increase in complexity, the spatial scales of cortical connections also grow, suggesting an evolution towards a global pattern of brain activities
\cite{RN27, RN35, RN1}.

Importantly, higher EPR, signifying increased energy consumption, contributes to better stability for memory states, enhancing the ability to retain information during delay periods. It appears that strengthening global connections yields a higher return on energy investment in terms of extended delay time. However, focusing solely on stability may not provide the complete picture, and the limitations of global connections remain uncertain. On one hand, it's essential to determine the extent to which extended delay times benefit individuals in real-life scenarios. On the other hand, as previous simulations have demonstrated, higher global connection strength leads to improved robustness
\cite{RN27}. Robustness encompasses the ability to maintain memory even when multiple areas in the cortex become inactive and the stability of attractors when facing disturbances. Nevertheless, alongside robustness, the network must possess mechanisms to release working memory. As global connections strengthen, more neural populations need to be suppressed to clear memory traces, potentially reducing flexibility. In summary, our findings providing quantification of thermodynamical cost suggest that, for functions at a similar level, a global pattern exhibits optimal energy efficiency, aligning with evolution with the assumption of optimization on energy efficiency. Further quantification of the benefits derived from robustness and flexibility may offer additional insights on the limitations in the future. 

\subsection{The dynamical and thermodynamical indicators for bifurcations and critical transitions of large-scale brain network}

   


\begin{figure}[htbp]
    \centering 
    \includegraphics[width=0.5\linewidth]{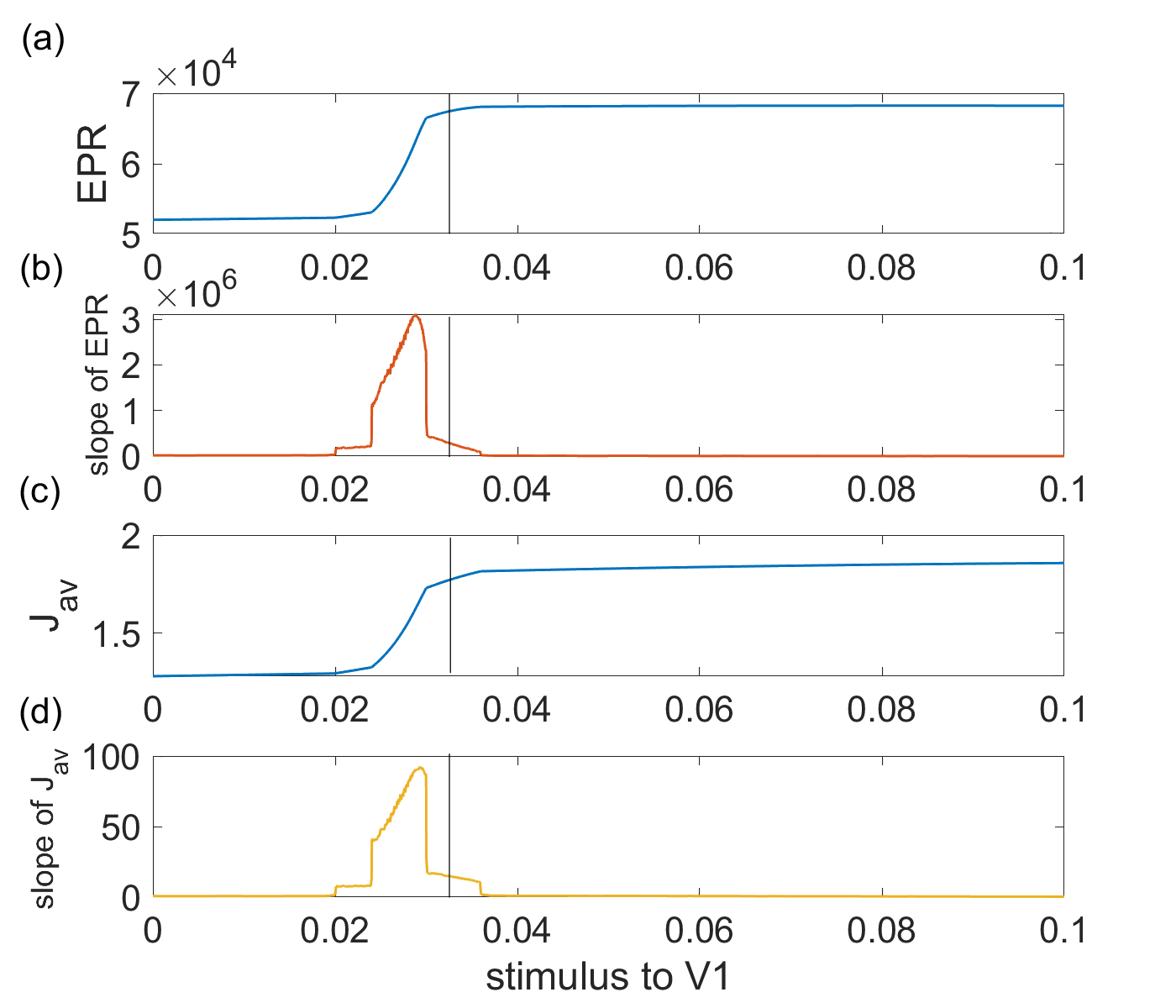}
    \caption{(a) The entropy production rate versus the stimulus intensity to primary visual cortex V1. (b) The slope of the entropy production rate. (c) The average flux strength versus the stimulus intensity to primary visual cortex V1. (d) The slope of the average flux strength.}
    \label{fig:fig4}
\end{figure}

\begin{figure}[htbp]
    \centering 
    \includegraphics[width=0.5\linewidth]{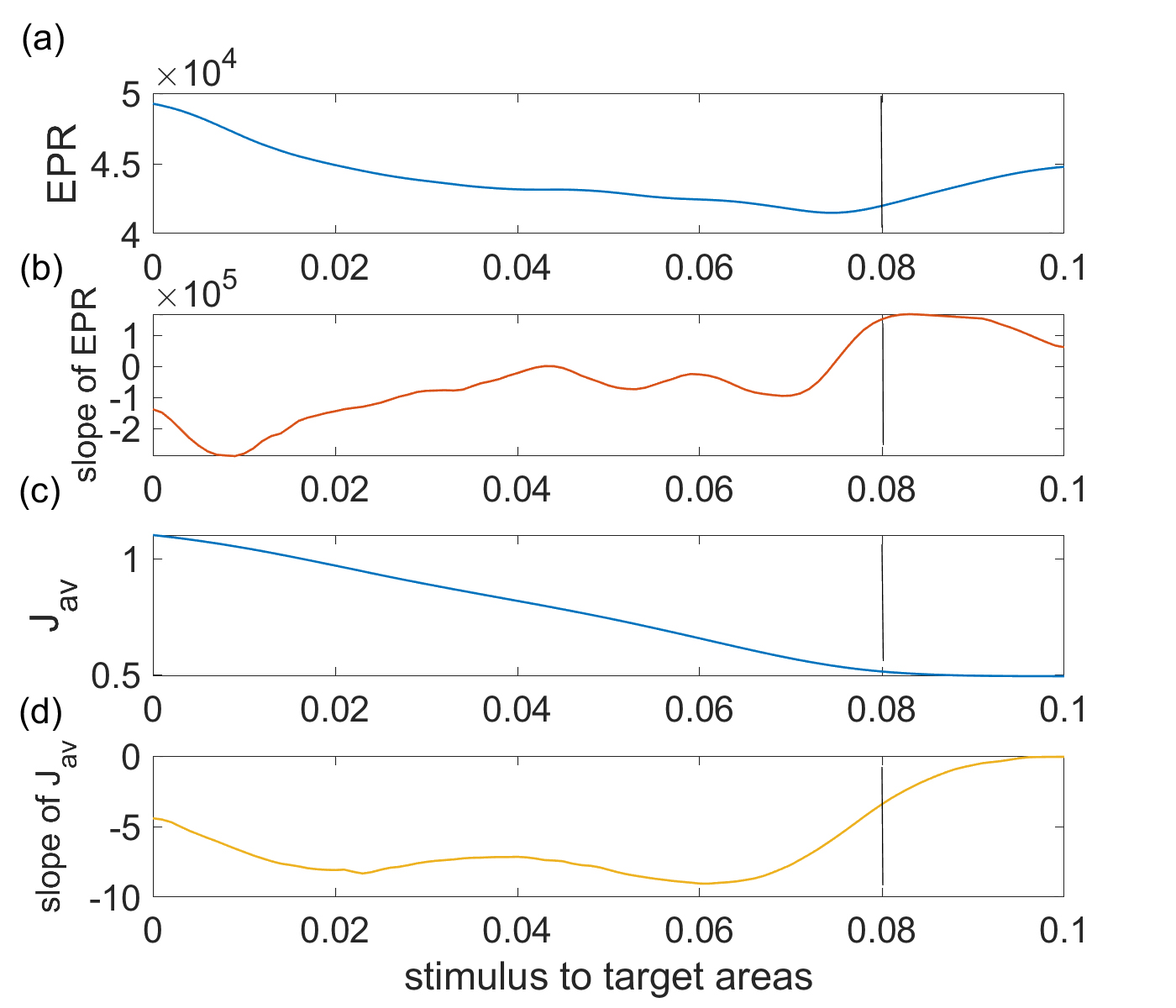}
    \caption{(a) The entropy production rate versus the stimulus intensity to inactivation target areas (9/46d, 9/46v, F7, 8B). (b) The slope of the entropy production rate. (c) The average flux strength versus the stimulus intensity to inactivation target areas (9/46d, 9/46v, F7, 8B). (d) The slope of the average flux strength.}
    \label{fig:fig5}
\end{figure}

Figure. 4 and Figure. 5 depicts the trend of the entropy production rate (EPR) and the average flux ($J_{\text{av}}$) versus stimulus intensity. As the stimulus intensity increases, both EPR and $J_{\text{av}}$ exhibit monotonic variations, with a notable change observed around input=0.034. The slope of EPR and $J_{\text{av}}$ clarifies the trend, revealing a peak at a relatively lower intensity than the bifurcation point. These trends are consistent for sensory activation and inactivation. In the case of increasing inputs to the inactivation target areas, the EPR and $J_{\text{av}}$ also demonstrate significant trends. Furthermore, as the stimulus intensity increases, both EPR and $J_{\text{av}}$ display a noteworthy increase, particularly during the saddle node bifurcation at the transition input=0.085. The nonzero flux drives the system away from equilibrium, resulting in the breakdown of detailed balance. During the state transitions, the dominant influence of non-equilibrium effects can be revealed through flux. While landscape tends to drive the system to the point attractor, the flux being rotational tends to be delocalized and thus destabilize the point attractor. This provides dynamical origin and valuable insights into the evolution of attractors and the emergence of bifurcations and phase transitions. Additionally, the flux is directly linked to the entropy production rate, offering a quantification of the thermodynamic cost and energy consumption at the system level during interactions. Hence, we anticipate that the rotational flux can characterize nonequilibrium effects that play a pivotal role in the emergence of new states, bifurcations, and critical transitions within the brain network. The dynamic and thermodynamic changes, as indicated by flux and entropy production, accompany the responding behaviors of the brain network during state transitions, suggesting their significance in understanding the origin of the state switching and potentially predicting the onset of critical transitions.



\begin{figure}[htbp]
    \centering
    \includegraphics[width=0.5\linewidth]{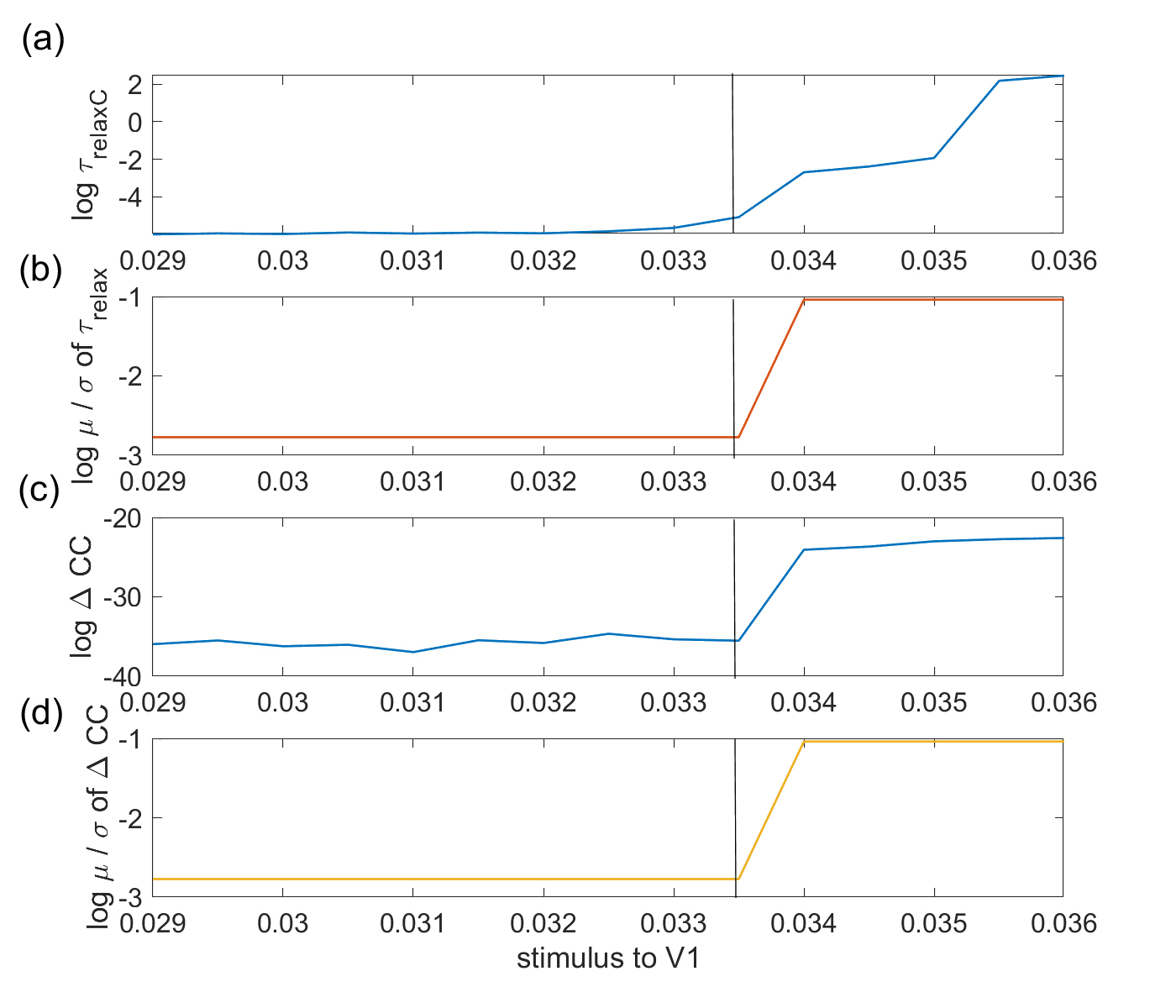}
    \caption{(a)(c) The trend of the relaxation period AutoC and difference in cross-correlations ($\Delta CC$) as the input strength (the stimulus intensity to primary visual cortex V1) approaches the critical. (b)(d) The variety of the ratio of mean to standard deviation of AutoC and $\Delta CC$ which are calculated from the differences in values between different subparts of the continuously selected trajectories.}
    \label{fig:fig6}
\end{figure}

\begin{figure}[htbp]
    \centering
    \includegraphics[width=0.5\linewidth]{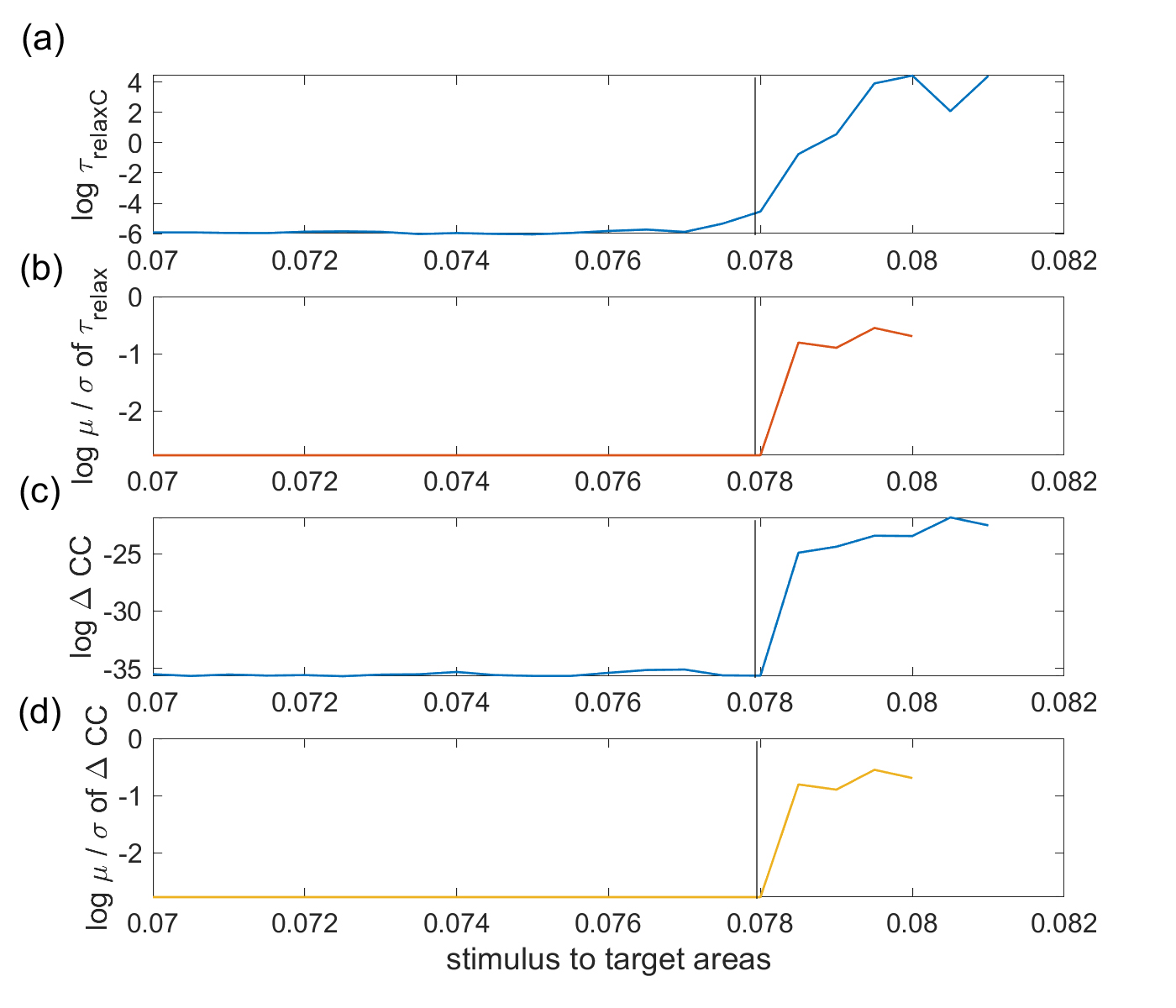}
    \caption{(a)(c) The trend of the relaxation period AutoC and difference in cross-correlations $\Delta CC$ as the input strength (the stimulus intensity to inactivation target areas (9/46d, 9/46v, F7, 8B)) approaches the critical. (b)(d) The variety of the ratio of mean to standard deviation of AutoC and $\Delta CC$ which are calculated from the differences in values between different subparts of the continuously selected trajectories.}
    \label{fig:fig7}
\end{figure}

The concept of critical slowing down theory suggests that during gradual changes in external environmental conditions, a system exhibits a phenomenon known as critical slowing down. This occurs as the system approaches a crucial bifurcation point (illustrated in Figure. 1a), characterized by a narrower basin of attraction and a flatter landscape (Figure. 1d, right attractor). For instance, the basin of attraction in the nonselective resting state, with an activation input value of $I_{\text{prb}}$ = 0.03, displays reduced width and depth, indicating critical slowing down. Furthermore, the nonselective basins in Figure. 1c become even smaller and shallower, providing further evidence of this behavior. During this phase, the system's "resilience" diminishes, increasing the risk of switching to an alternative stable state 
\cite{RN62}. To capture these tipping points characterized by saddle-node bifurcations, critical slowing down has been proposed as a useful approach, leading to slower recovery from perturbations, increased autocorrelation, and higher variance
\cite{RN62}. After an equilibrium, the system experiences an external transient disturbance, it eventually returns to its original state, a process known as relaxation. The relaxation time represents the period required for the system to adapt to environmental changes. In our study, we depict the relaxation time based on autocorrelation (Figure. 6a, Figure. 7a). A significant increase in relaxation time is observed near the bifurcation point, specifically $I_{pulse}$ = 0.035 in the activation process and $I_{\text{prb}}$ = 0.085 in the inactivation process. Close to these bifurcations, the system requires considerably more time to revert to its initial equilibrium state following external perturbation. 

Moreover, Figure. 6c and Figure. 7c illustrate the average difference in cross-correlations ($\Delta CC$) between forward and backward time, serving as a measure of time irreversibility. At equilibrium, $C_{XY} (\tau)$ and $C_{YX} (\tau)$ are identical, indicating time reversibility of the system 
\cite{RN40}. The degree of detailed balance breaking 
\cite{RN63, RN64} is directly linked to time irreversibility. Hence, the difference in cross-correlations quantifies both the degree of nonequilibriumness and the strength of the flux. Although directly quantifying flux from experimental data is challenging, assessing the difference in cross-correlations between forward and backward time for observed real-time traces of observables provides a practical means to quantify the nonequilibrium driving force of the system. Notably, $\Delta CC$ exhibits a significant increase near the regime of bifurcation, where the resting state or the memory state vanishes. Additionally, when the sensory stimulus increases, the change in $\Delta CC$ occurs well before the bifurcation, indicating the disappearance of the resting state at $I_{input}$ = 0.035. Similarly, when the system undergoes inactivation inputs on target areas, the indicator responds at an early stage. This practical method enables the early detection of warning signals for the temporal trajectory of the multiarea brain network system, providing valuable insights into the onset of state transitions. 

Furthermore, both the relaxation time and the difference in cross-correlations exhibit a trend of early increase. To further illustrate the observed changes in autocorrelation (AutoC) and the difference in cross-correlations ($\Delta CC$), we plot variety of the ratios of mean to standard deviation of both AutoC and $\Delta CC$ (Figure.6d, Figure.7d). The mean is calculated from the entire test trajectories, while the standard deviation is computed from the differences in values between different subparts of continuously selected trajectories. These ratios convey the variety and the fluctuations of both characteristics. A longer resilience time corresponds to more detail balance being broken, with a longer period remaining before returning to equilibrium. Consequently, trajectories during periods shorter than the resilience time present high irreversibility or symmetry broken. Accordingly, the values of AutoC and $\Delta CC$ during sub-resilience time are also unstable, and the ratios exhibit more obvious trends than mean values. These early warning signals can be utilized to detect emerging state transitions in our brain network. Moreover, within the intricate landscape of the brain network dynamics, the states and transitions may not always manifest themselves prominently . They may remain concealed, particularly in cases involving organic disorders. Given that the detection of critical slowdown and time irreversibility can be readily achieved through the analysis of time series data and holds a well-defined physical significance, this approach can prove useful in uncovering latent state transitions. These concealed transitions could potentially offer valuable insights into neural mechanisms and provide support for clinical investigations.  

\section{Materials and Methods}

\subsection{Working memory in a large-scale network of macaque neocortex with statistical fluctuations.}

An anatomically constrained computational model of large-scale macaque cortex was developed to elucidate the circuit mechanism of distributed working memory. The model has been comprehensively described and discussed in Wang et al.'s recent article
\cite{RN27}. In this section, we will provide a brief overview of the model and introduce a modification aimed at enhancing its dynamical and thermodynamic rationality. The model performs the interactions among the local intra-areal scale and the global inter-areal scale. The Wong-Wang model
\cite{RN14} was employed to describe the neural dynamics of the local microcircuit representing a cortical area. This model, in its three-variable version, captures the temporal evolution of firing rates in two input-selective excitatory populations and an inhibitory population, all interconnected. The model is governed by the following equations: 
$$\frac{\text{dS}_A}{\text{dt}}=-\frac{S_A}{\tau _N}+\gamma  \left(1-S_A\right) r_A+S_{\text{noise}}$$
$$\frac{\text{dS}_B}{\text{dt}}=-\frac{S_B}{\tau _N}+\gamma  \left(1-S_B\right) r_B+S_{\text{noise}}$$
$$\frac{\text{dS}_C}{\text{dt}}=-\frac{S_C}{\tau _G}+\gamma _i r_C+S_{\text{noise}}$$
In the above equations, $S_A$ and $S_B$ represent the NMDA conductances of the selective excitatory populations A and B, respectively, while $S_C$ denotes the GABAergic conductance of the inhibitory population. Here we have introduced a set of stochastic force terms, referred to as $S_{noise}$, which have been generated through Gaussian white noise. The motivation behind this introduction lies in the fact that the dynamics of each local circuit are acquired through the utilization of the mean-field method. The unavoidable thermodynamic fluctuations and the finite robustness of neurons inherently introduce elements of uncertainty. It is noteworthy that this noise is of small magnitude and typically becomes submerged within the background noise originated from synaptic currents in the simulations. This form of additive noise does not exert influence on the overall dynamics of the system as a whole; its effects only manifest in specific populations with exceedingly weak input current conditions. The constants are assigned the values $\tau_N$ = 60 ms, $\tau_G$ = 5 ms, $\gamma$ = 1.282, and $\gamma_I$ = 2. The mean firing rates of the excitatory populations ($r_A$ and $r_B$) and the inhibitory population ($r_C$) are obtained by solving the transcendental equation $r_i=\phi _i (I_i )$ at each time step. Here, $I_i$ represents the input to population 'i', given by the following equations: 
$$I_A=J_S S_A + J_C S_B+J_{\text{EI}} S_C +I_{A,\text{net}}+I_{0 A}+I_{\text{noise}} (t)$$
$$I_B=J_C S_A+ J_S S_B+ J_{\text{EI}} S_C+I_{B,\text{net}}+ I_{0 B}+I_{\text{noise}} (t)$$
$$I_C=J_{\text{IE}} S_A+ J_{\text{IE}} S_B+ J_{\text{II}} S_C+I_{C,\text{net}}+ I_{0 C}+I_{\text{noise}} (t)$$
In the above expressions, $J_S$ and $J_C$ represent the self- and cross-coupling strengths between excitatory populations, $J_{\text{EI}}$ signifies the coupling from inhibitory populations to any excitatory population, $J_{\text{IE}}$ denotes the coupling from any excitatory population to the inhibitory population, and $J_{\text{II}}$ represents the self-coupling strength of the inhibitory population. The parameters $I_{0i}$ (i = A, B, C) correspond to the background inputs of each population. Additionally, to introduce another stochasticity into the syste, $I_{noise}$ is introduced as a Gaussian white noise. The specific parameter values are $J_S$ = 0.3213 nA, $J_C$ = 0.0107 nA, $J_{\text{IE}}$ = 0.15 nA, $J_{\text{EI}}$ = -0.31 nA, $J_{\text{II}}$ = -0.12 nA, $I_{0 A}$ = $I_{0 B}$ = 0.3294 nA, and $I_{0C}$ = 0.26 nA.
The transfer function $\phi _i(I_i)$ transforms the input into firing rates for the excitatory populations and takes the form:
$$ \phi _{A,B} (I)=\frac{a I-b}{1-\exp (-d (a I-b))}$$
For the inhibitory population, we use a threshold-linear function:
$$ \phi _C (I)=\left[\frac{1}{\left(c_1 I-c_0\right) g_I}+r_0\right]_+$$
In the above equation, $[x]_+$denotes rectification. The parameter values for the excitatory populations are $a$ = 135 Hz/nA, $b$ = 54 Hz, and $d$ = 0.308 s, while the inhibitory population is characterized by $g_I$ = 4, $c_1$ = 615 Hz/nA, $c_0$ = 177 Hz, and $r_0$ = 5.5 Hz.

Before considering the large-scale network and inter-areal connections, area-to-area heterogeneity was incorporated into the model. The cortical system consists of $N$ = 30 local cortical areas, with available inter-areal connectivity data. Each cortical area is represented as a Wong-Wang model similar to the one described earlier. However, rather than assuming all areas to be identical, natural area-to-area heterogeneity observed in anatomical studies was accounted for. 

To introduce heterogeneity, the gradient of dendritic spine density across cortical areas was considered, ranging from low spine numbers ($\sim$600) in early sensory areas to high counts ($\sim$9000) in higher cognitive areas. Additionally, the anatomical hierarchy of cortical areas, where the hierarchical position of an area 'i' ($h_i$) is computed using a generalized linear model based on supra-granular layer neuron fractions (SLN) projecting to and from that area.

In this approach, hierarchical values ($h_i$) are assigned to each area, such that the difference in values predicts the SLN of a projection. The hierarchical values are then normalized as $h_i=h_i/h_{\max }$. By using this regression-based approach, a gradient of values is created for each area, which correlates with dendritic spine counts. 

In the subsequent analysis, the synaptic strength (both local and long-range) of each area is determined as a linear function of the observed dendritic spine counts or their corresponding hierarchical positions. The resulting large-scale network exhibits a gradient of local and long-range recurrent strength, with sensory areas having weaker local connectivity compared to association areas. The local and long-range strength value for a given area 'i' in this gradient is denoted as $h_i$, ranging from 0 (bottom of the gradient, area V1) to 1. 

To summarize, the self-coupling strength $J_S (i)$ follows a gradient, with its value varying from $J_{\min}$ to $J_{\max }$. The formula  $J (i)=h_i \left(J_{\max }-J_{\min }\right)+J_{\min }$ captures the gradient of synaptic strengths. Assuming a fixed spontaneous activity level for all areas ensures that the spontaneous firing rate remains physiologically realistic throughout the model. This is achieved by maintaining the quantity $J_C+2 \zeta  J_{\text{EI}} J_{\text{IE}}+J_S\equiv J_0$ constant, where $J_0$ is 0.2112 nA for the original parameter values described earlier. Additionally, the coupling from excitatory to inhibitory neurons, $J_{IE}$, scales with the ranks and with $J_S$ according to $J_{\text{IE}}= J_{\text{EI}} \zeta \left(J_0-J_C-J_S\right)/2$, thereby ensuring a consistent spontaneous solution across all areas. 

By implementing these synaptic strength gradients, the model encompasses distributed working memory (WM) without inherently bistable areas, assuming $J_{\min}$ = 0.21 nA and $J_{\max }$ = 0.42 nA (below the critical value). This configuration enables sustained activity in the model as a result of global cooperative effects arising from inter-areal interactions, rather than intrinsic bistability.

Next, the inter-areal projections connect the isolated cortical areas to form the large-scale network. These projections originate exclusively from excitatory neurons, as inhibitory projections tend to be localized within circuits. The long-range input term received by each population within an area '$x$' from all other cortical areas can be expressed as follows: 
$$I_{A,\text{net}}^\text{x}=G \sum _y W_{\text{xy}} \text{SLN}_{\text{xy}} S_{\text{Ay}}$$
$$I_{B,\text{net}}^\text{x}=G \sum _y W_{\text{xy}} \text{SLN}_{\text{xy}} S_{\text{By}}$$
$$I_{C,\text{net}}^\text{x}=\frac{G}{Z}\sum _y W_{\text{xy}}\left(1-\text{SLN}_{\text{xy}}\right)\left(S_{\text{Ay}}+S_{\text{By}}\right)$$
In these equations, $G$ represents the global coupling strength, $Z$ is a balancing factor, and $W$ denotes the connectivity matrix (further details provided below). The superscripts denote the cortical area, while the subscripts indicate the specific populations within each area. The sums in above equations encompass all cortical areas in the network ($N$ = 30). Excitatory populations A and B receive long-range inputs from equally selective units in other areas, while inhibitory populations receive inputs from both excitatory populations. This means that population A in a given area can be influenced directly by A-selective neurons from other areas and indirectly by B-selective neurons from other areas through local interneurons.

The global coupling strength, $G$, controls the overall strength of long-range projections in the network (default value: $G$ = 0.48, unless stated otherwise). The balancing factor $Z$ determines the relative balance between long-range excitatory and inhibitory projections, with $Z$ = 1 indicating equal strengths for both types. Achieving actual balance within the target area requires considering local connections 
\cite{RN23}. For inhibitory populations with linear transfer functions, $Z$ is calculated as follows: 
$$Z=2c_l\tau G\gamma _IJ_{\text{EI}}/\left[\gamma _I G \tau  c_l J_{\text{II}}-g_I\right]$$
Apart from global scaling factors, the effect of long-range projections from population '$y$' to population '$x$' is influenced by two factors. The first factor, $W_{xy}$, represents the anatomical projection strength obtained from tract-tracing data 
\cite{RN65}. They rescale these strengths to align them with anatomical data by maintaining the proportions between projection strengths. The rescaling equation is as follows:
$$W_{\text{xy}}=k_1\left(\text{FLN}_{\text{xy}}\right)^{k_2} $$
In this equation, $k_1$ = 1.2 and $k_2$ = 0.3. The same qualitative behavior can be achieved using different parameter values or rescaling functions, as long as the network is set within a standard working regime, allowing signals to propagate across areas while avoiding global synchronization. The FLN values are normalized so that $\sum _y \text{FLN}_{\text{xy}}=1$. This normalization ensures better control over the heterogeneity levels of each area and minimizes confounding factors such as uncertainty in tract tracing experiment injections and potential influences of homeostatic mechanisms.

Similar to local connections, they introduce a gradient of long-range projection strengths using spine count data
\cite{RN27}. Specifically, $W_{xy}$ is multiplied by ($J_S (x)/J_{\max }$) to align the long-range projections with the same gradient observed in local connectivity. Another factor to consider is the directionality of signal propagation across the cortical hierarchy. It is assumed that feedforward (FF) projections are predominantly excitatory, facilitating signal transmission from sensory to higher areas. Conversely, feedback (FB) projections with a preferential inhibitory nature contribute to the emergence of realistic distributed working memory (WM) patterns. Gradually introducing this feature, this links the inter-areal projections to the SLN data, which serves as a proxy for the FF/FB nature of a projection ($SLN$ = 1 represents purely FF, while $SLN$ = 0 represents purely FB). In the model, they assume a linear dependence on SLN for projections to excitatory populations and (1 - $SLN$) for projections to inhibitory populations, as shown earlier.

\subsection{Landscape and flux theory for large-scale brain network}

In diverse natural systems, stochastic fluctuations play a significant role
\cite{RN39, RN67, RN38, RN40, RN66}. To describe the nonlinear dynamics influenced by random fluctuations, we introduce the following equation:
$$\dot{x}=F (x)+\eta (x,t)$$
Here, $F(x)$ denotes the deterministic force, and the state of the system is represented by the vector $x$, $x=[S_i]$ donates the conductances in this work. Stochastic dynamic equation with multiplicative noise. The force $\eta (x,t)$  corresponds to Gaussian fluctuating force with an autocorrelation function given as $<\eta (x,t),\eta (x,0)>=2 D (x) \delta (t)$, where $D(x)$  is the diffusion coefficient matrix. Set $D(x)=DG(x)$, where $D$ is the diffusion coefficient representing the noise intensity and $G$ is the scaled diffusion matrix describing the anisotropy. 

For the large scale distributed working memory model, the indeterministic part is as a noise input component of the synaptic current. As the noise intensity is quite low comparing with the major components of synaptic stimulus, we apply linear expansion to the responding function $\phi(x)$. 
$$\dot{S}_{A,B}=-\frac{S_{A,B}}{\tau _n}+\gamma  \phi   \left(1-S_{A,B}\right) \left(I_i+I_{\text{noise}}\right)=-\frac{S_{A,B}}{\tau _n}+\gamma (1-S_{A,B})\left( \phi (I_i) +\frac{\partial \phi }{\partial I}|_{I_i}I_{\text{noise}}\right),$$
$$\dot{S}_C=-\frac{S_C}{\tau _g}+\gamma _I \phi _C \left(I_i+I_{\text{noise}}\right)=-\frac{S_C}{\tau _g}+\gamma _I\left( \phi _C (I_i)+\frac{\partial \phi _C}{\partial I}|_{I_i}I_{\text{noise}}\right).$$ Consider $I_{\text{noise}}$ has a form of Gaussian white noise, the anisotropic diffusion coefficient matrix can be written thus
$$D_{ \text{ii} A,B}=\frac{1}{2 \tau }\gamma ^2 \sigma ^2 \left(1-x_i\right){}^2 \left( \phi '\left(I_i (x)\right)\right){}^2,$$
$$D_{\text{ii} C}=\frac{1}{2 \tau }\gamma _I^2 \sigma ^2 \left(\phi _C'\left(I_i (x)\right) \right){}^2,$$
$$D_{i\neq j}=0.$$
The evolution of probability density function of the system, $P(x,t)$, obeys the local conservation law: $\frac{\partial P}{\partial t}=-\nabla \cdot J$. The change of the probability in time is equal to the net flux $J$ in or out. The probability flux $J$ is defined as: $J(x,t)=F(x)P(x,t)-\nabla (D(x)P(x,t))$. The driving force for the dynamics can then be decomposed as: $F=-D G\cdot \nabla U+J_{\text{ss}}/P_{\text{ss}}+D\nabla \cdot G$, where $U=-\ln  P_{\text{ss}}$ is the nonequilibrium potential landscape which is related to the steady-state probability distribution $(P_{ss})$ and the steady-state probability flux $J_{ss}$ 
\cite{RN69, RN68, RN38}. 

For system having the curl nature at nonequilibrium steady-state $\nabla \cdot J_{\text{ss}}=0$, the nonzero probability flux is $J_{\text{ss}}=F P_{\text{ss}}-\nabla \left(D P_{\text{ss}}\right)=F P_{\text{ss}}-D\cdot \nabla P_{\text{ss}}-(\nabla \cdot D)P_{\text{ss}}$, which donates the net flow in or out of the system, measuring the degree of detail balance breaking away from equilibrium. 

To comprehend the underlying potential landscape $U$ defined as $-ln(P_{ss})$, the initial step involves calculating the probability distribution of the steady state. However, solving the corresponding diffusion equation directly becomes impractical due to its high dimensionality. To overcome this challenge, we adopt the self-consistent mean-field approximation, which simplifies the complexity by breaking down the probability distribution into products of individual probabilities. This approach allows each neuron to interact with an average field from other neurons, effectively reducing the degrees of freedom and making the problem computationally feasible. 

In practice, we employ moment equations as a more accessible means to approximate the behavior of neural networks
\cite{RN70, RN49, RN37}. By assuming a Gaussian distribution and focusing on scenarios with a small diffusion coefficient ($D$), we arrive at simplified moment equations for the mean ($u(t)$) and variance ($\Sigma (t)$) of the neural network: 
$$\dot{u}(t)=C[u(t)]$$
$$\dot{\Sigma }(t)=\Sigma  (t) A^T+A \Sigma  (t)+2 D[u(t)]$$
In these equations, $u(t)$, $\Sigma (t)$, and $A(t)$ represent vectors and tensors denoting the mean, variance, and transformation matrix, respectively. By solving these equations, we acquire the values for $u(t)$ and $\Sigma (t)$ that describe the neural network's evolution. Furthermore, in a multi-stable system, not all point stable states carry the same weight, indicating that some states are more probable than others. To establish the relative weights of these attractors, we analyze the percentage of each final attractor state originating from different initial conditions. This enables us to determine the weight coefficients of each stable state. 

In accordance with the mean-field approximation, the steady-state probability distribution is represented as $P_{\text{ss}}^l=\frac{1}{\sqrt{(2 \pi )^n \left| \Sigma ^l\right| }}e^{-\frac{(x-\mu ^l) (x-\mu ^l)^T}{2 \Sigma ^l}}$, where $\mu$ denotes the mean of the state vector near the $l$th stable point or attractor, and $\Sigma ^l$ corresponds to the variance matrix of the probability distribution. The total distribution is expressed as $P_{\text{ss}}=\underset{l}{\Sigma } w_l P_{\text{ss}}^l$, with $w_l$ representing the weight of the $l$th attractor. 

For state near $x=[x_{fix}]^l$, the driving force can be presented as $F (x)=A \left(x-x_{\text{fix}}\right)$, where $A$ is the Jacobi matrix of $F(x)$ at $[x_{fix}]^l$. Also, we can simplify the flux distribution as 
$J_l=A_l \left(x-x_{\text{fix},l}\right)P_l +\frac{D_0}{\Sigma }  \left(x-x_{\text{fix},l}\right)P_l+\nabla \cdot D|_{x=x_{\text{fix}}}P_l=\left(M \left(x-x_{\text{fix},l}\right)+N\right)P_l.$
Set $M=A+D_0/\Sigma$, $N=-\nabla \cdot D|_{x=\text{xfix}}.$
The divergence of $D(x)$ is given:
$$\frac{\partial D_{\text{ii} A,B}}{\partial x_i}=\frac{\gamma ^2 \sigma ^2}{\tau } \left(1-x_i\right){}^2  J_s \phi '\left(I_i (x)\right) \phi ''\left(I_i (x)\right)-\frac{\gamma ^2 \sigma ^2}{\tau } \left(1-x_i\right) \left( \phi '\left(I_i (x)\right)\right){}^2,$$
$$\frac{\partial D_{ \text{ii} C}}{\partial x_i}=\frac{\gamma _I^2 \sigma ^2 }{\tau } J_{\text{II}} \phi _C'\left(I_i (x)\right) \phi _C''\left(I_i (x)\right).$$

\subsection{Nonequilibrium Thermodynamics}

Non-equilibrium open systems engage in exchanges of energy, materials, and information with their surroundings. The time evolution of the system's entropy can be broken down into two components: the entropy production rate and the heat dissipation rate 
\cite{RN73, RN71, RN72, RN56, RN74}: 
$$\dot{S}=\dot{S}_t-\dot{S}_e$$
The population entropy production rate can be represented as $$e_p=\dot{S}_t=\int \text{dx}  \left( J\cdot (D G)^{-1}\cdot J\right)/{P},$$ and $\dot{S}_e=\int \text{dx}\left(J\cdot (D G)^{-1}\cdot (F-D\nabla \cdot G)\right)$ denotes the heat dissipation rate of the environment. Hence, the entropy production rate can be viewed as the total entropy change of the system and its environment: $e_p=\dot{S}_t=\dot{S}+\dot{S}_e$. It is essential for the entropy production rate to be non-negative, indicating that it must not be negative, while the heat dissipation rate may be either positive or negative. This measure enables the quantification of entropy flow rate between the environment and the non-equilibrium system. At steady-state, the entropy production rate and the heat dissipation rate are equal 
\cite{RN73, RN71, RN56, RN74}, providing a global thermodynamic characterization of the non-equilibrium system. Additionally, we define the average magnitude of the flux as $\overset{-}{J}=\int \left|J^2\right|\text{dx}$ to gauge the extent to which a system deviates from equilibrium.

\subsection{Quantification of dominant path}

The utilization of the path integration method allows for the identification and quantification of the most probable transitions occurring between two stable states. The path integration expression that characterizes the likelihood of a trajectory from the initial state $x_i$ at time $t$ = 0 to the final state $x_f$ at time $t$ is represented by equation 
\cite{RN75} $$P\left(x_f,t_f\left|x_i\right.,0\right)=\int \delta\text{x}\exp \left[-\int \text{dt}\left(\frac{1}{2}\nabla \cdot F(x)+\frac{1}{4} \left(\frac{\text{d}x}{\text{dt}}-F (x)\right)\cdot (D G)^{-1}\cdot \left(\frac{\text{d}x}{\text{dt}}-F (x)\right)\right)\right]$$$$=\int \delta\text{x} \exp (-A (x))=\int \delta\text{x} \exp (-\int  L (x(\text{t}))\text{dt}),$$ where $L (x(\text{t}))$ stands for the Lagrangian and $A(x)$ represents the action corresponding to each trajectory across the potential landscapes. The integration over $\delta\text{x}$ encompasses a summation over all possible trajectories connecting $x_i$ at time 0 to $x_f$ at time $t_f$. The exponential term assigns weight to each specific path, and therefore, the probability of transitioning from $x_i$ to $x_f$ involves aggregating the weighted sum of all conceivable paths. The path integration can be approximated by the trajectory that contributes the most to the weight, given that the contributions of other paths dwindle exponentially. The most influential paths, along with their optimal weights, can be ascertained through the minimization of the action $A(x)$ or the Lagrangian $L (x(\text{t}))$, since the probability of the dominant path is proportional to $exp[-A(x)]$. Additionally, when choosing a large enough time scale as the integral interval of the action, the results appear to be insensitive to the upper limit of the time integral $t_f$. Consequently, the paths that exert the most substantial influence on the weight are identifiable as the prevailing routes governing transitions between savanna and forest ecosystems. The underpinning of the path integration framework here relies on the Onsager–Machlup functional for a diffusion process in the presence of finite fluctuations 
\cite{RN76}. In the scenario of negligible noise, the divergence of the force component in the Onsager–Machlup functional can be disregarded. This simplification leads to a form of the path integration framework akin to that originating from the Freidlin–Wentzell theory.

\section{Conclusion}

In this paper, we delve into the historical exploration of cognitive behaviors and introduce the application of statistical physics mean-field theory in modeling neuron population spiking and interactions. Although the previous model effectively illustrates the dynamic processing of working memory, it falls short in providing a physical explanation for the crucial role of specific structural settings, particularly concerning nonsymmetric inter-area connections and connection distances. To fill in this gap, we quantify the potential landscape of the large-scale network, highlighting the attractors and diffusion in the state space. Building upon this analysis, we focus on measuring the dominant paths between memory and resting states, revealing the most probable transition paths within the network during cognitive tasks. Our examination of the dynamical driving force and nonequilibrium diffusion reveals distinct forward and backward paths, indicating a specific temporal order among hierarchical areas, which sheds light on the intrinsic dynamics disparities in cortical regions stemming from nonsymmetric connections.

As a living system, the exchange of material, energy, and information between the neural network and the environment is a continual process sustaining its activities. Within the realm of thermodynamics, the energy exchange is manifested in the entropy flow between the system and the environment. We quantify the variations of entropy production rates concerning local and global connection strengths in the large-scale neural network's nonequilibrium steady state, thereby measuring the thermodynamic costs of maintaining working memory function. Simultaneously, we assess the robustness of memory sustainability through the lens of relative stability, mean first passage time and action. Our investigation into the relationship between functional features and connection patterns indicates that increasing connection strength contributes to heightened stability and entropy production rates. However, a higher ratio of global connections results in lower entropy production rates while maintaining the same level of stability. Thus, the limitation and optimization of the entropy production emerges as a potential trigger for the increased connection distance observed in the mammalian brain. 

Furthermore, the stimulation of sensory areas or inactivation prompts state switching within the cortical network, accompanied by characteristic variations in both thermodynamics and dynamics. To capture these changes, we calculate entropy production rates and average probability flux strength, quantifying time irreversibility and critical slowdown. These quantitative characteristics exhibit specific trends preceding bifurcations or catastrophes, offering early detection or warning capabilities. Notably, irreversibility and critical slowdown are readily accessible in experimental settings.  

\section*{Acknowledgement}

X.W. thanks the support by the National Natural Science Foundation of China Grant No. 21721003 and No.12234019. We thank the helpful discussion with Dr. Han Yan, Dr. Li Xu and Dr. Liufang Xu.

\section*{Data Availability}

All data is included in the manuscript and/or supporting information.

\bibliographystyle{apsrev4-2}
\bibliography{LSWM}  

\end{document}